\documentclass[11pt]{article}

\usepackage{graphicx,amsmath,amsfonts,amssymb,dcolumn,amsthm,euscript,braket,xcolor,pslatex,slashed}

\numberwithin{equation}{section}

\newcommand{\p}{\partial}
\newcommand{\ophi}{\overline\varphi}

\definecolor{CraneRed}{rgb}{0.99,0.20,0.02}

\newcommand{\MSbar}{\overline{\mbox{MS}}}

\begin{document}

\title{\textbf{ Non-perturbative aspects of Euclidean Yang-Mills theories in linear covariant gauges: Nielsen identities and a BRST invariant two-point correlation function}}
\author{\textbf{M.~A.~L.~Capri$^{a}$}\thanks{caprimarcio@gmail.com},
\textbf{D.~Dudal$^{b,c}$}\thanks{david.dudal@kuleuven.be},
\textbf{A.~D.~Pereira$^{a}$}\thanks{duarte763@gmail.com},
\textbf{D.~Fiorentini$^{a}$} \thanks{diegofiorentinia@gmail.com},
\textbf{M.~S.~Guimaraes$^{a}$}
\thanks{msguimaraes@uerj.br}, \\
\textbf{B.~W.~Mintz$^{a}$}\thanks{bruno.mintz@uerj.br},
\textbf{L.~F.~Palhares$^{a}$} \thanks{leticia.palhares@uerj.br},
\textbf{S.~P.~Sorella$^{a}$}\thanks{silvio.sorella@gmail.com}, \\\\\
\textit{{\small $^{a}$ UERJ -- Universidade do Estado do Rio de Janeiro,}}\\
\textit{{\small Instituto de F\'isica -- Departamento de F\'{\i}sica Te\'orica -- Rua S\~ao Francisco Xavier 524,}}\\
\textit{{\small 20550-013, Maracan\~a, Rio de Janeiro, Brasil}}\\
\textit{{\small $^b$ KU Leuven Campus Kortrijk -- KULAK, Department of Physics, Etienne Sabbelaan 51 bus 7800,}}\\
\textit{{\small 8500 Kortrijk, Belgium}}\\
\textit{{\small $^c$           Ghent University, Department of Physics and Astronomy, Krijgslaan 281-S9, 9000 Gent, Belgium}}}
\date{}
\maketitle

\begin{abstract}
In order to construct a gauge invariant two-point function in a Yang-Mills theory, we propose the use of the all-order gauge invariant transverse configurations $A^h$. Such configurations can be obtained through the minimization of the functional $A^2_{min}$ along the gauge orbit within the BRST invariant formulation of the Gribov-Zwanziger framework recently put forward in \cite{Capri:2015ixa,Capri:2016aqq} for the class of the linear covariant gauges. This correlator turns out to provide a characterization of  non-perturbative aspects of the theory in a BRST invariant and gauge parameter independent way. In particular, it turns out that the poles of $\langle A^h_\mu(k) A^h_\nu(-k) \rangle$ are the same as those of the transverse part of the gluon propagator, which are also formally shown to be independent of the gauge parameter $\alpha$ entering the gauge condition through the Nielsen identities. The latter follow from the new exact BRST invariant formulation introduced before. Moreover, the correlator  $\langle A^h_\mu(k) A^h_\nu(-k) \rangle$ enables us to attach a BRST invariant  meaning to  the possible positivity violation of the corresponding temporal  Schwinger correlator, giving thus for the first time a consistent, gauge parameter independent, setup to adopt the  positivity violation of $\langle A^h_\mu(k) A^h_\nu(-k) \rangle$ as a signature for gluon confinement. Finally, in the context of gauge theories supplemented with a fundamental Higgs field, we use $\langle A^h_\mu(k) A^h_\nu(-k) \rangle$ to probe the pole structure of the massive gauge boson in a gauge invariant fashion.
\end{abstract}

\section{Introduction}
In this paper we give a sequel to our previous works \cite{Capri:2015ixa,Capri:2016aqq}, where an exact BRST invariant local formulation for the Gribov-Zwanziger (GZ) framework \cite{Gribov:1977wm,Zwanziger:1988jt} was derived in the class of the linear covariant gauges. In its original original version \cite{Gribov:1977wm,Zwanziger:1988jt}, the Gribov-Zwanziger setup was outlined in the Landau gauge, $\partial_\mu A^a_\mu=0$, in order to take into account the non-perturbative phenomenon of the existence of Gribov copies, which affects the Faddeev-Popov quantization prescription.

According to \cite{Gribov:1977wm,Zwanziger:1988jt}, the main idea to face the issue of the Gribov copies was to restrict  the functional integral to a certain region $\Omega$ in field space, called the Gribov region, which is defined as
\begin{equation}
\Omega = \{ \; A^a_\mu|  \;  \partial_\mu A^a_\mu =0,  \;    {\mathcal M}^{ab}(A) > 0 \; \}    \;, \label{om}
\end{equation}
where ${\mathcal{M}}^{ab}(A)$ is the Hermitian Faddeev-Popov operator
\begin{equation}
\mathcal{M}^{ab}=-\delta^{ab}\partial^2+gf^{abc}A^{c}_{\mu}\partial_{\mu},\,\,\,\, \mathrm{with}\,\,\,\, \partial_{\mu}A^{a}_{\mu}=0\,.
\label{intro0}
\end{equation}
Later on, important properties of the region $\Omega$ were rigorously established \cite{Dell'Antonio:1989jn}, namely:
\begin{itemize}

\item[\it i)] $\Omega$ is  convex, a property which follows from the linearity of the Faddeev-Popov operator ${\mathcal{M}}^{ab}$.

\item[\it ii)] $\Omega$ is bounded in all directions in field space. The boundary $\partial \Omega$, where the first vanishing eigenvalue of the Faddeev-Popov operator shows up,  is called the first Gribov horizon.

\item[\it iii)] Every gauge orbit crosses at least once the region $\Omega$.
\end{itemize}
In particular, property {\it iii)} gives a well defined support to the restriction to the region $\Omega$. Remarkably, a local and renormalizable action\footnote{We remind here that $\Omega$ itself  is not completely free from Gribov copies \cite{vanBaal:1991zw,semenov},  i.e.~additional copies still exist inside $\Omega$.   A smaller region within $\Omega$ exists which is fully free from Gribov copies. This region is called  the fundamental modular region. Though, unlike the case of  the Gribov region $\Omega$, a local and renormalizable framework implementing the restriction to the fundamental modular region is, at present,  unknown. Therefore, we shall proceed by focusing on the region $\Omega$.} can be constructed for the restriction to $\Omega$: the so-called Gribov-Zwanziger action, see \cite{Vandersickel:2012tz} for a general review.

In \cite{Capri:2015ixa,Capri:2016aqq}, we have been able to move away from the Landau gauge,  generalizing the Gribov-Zwanziger construction to the class of the linear covariant gauges, i.e.~$\partial_\mu A_\mu= i\alpha b$, where $\alpha$ is the (non-negative) gauge parameter. Obviously, the Landau gauge can be seen as a particular case of the linear covariant gauges, corresponding to $\alpha=0$. Moreover, as already mentioned, we were able, for the first time,  to write down an exact nilpotent BRST symmetry of the Gribov-Zwanziger action in the linear covariant gauges which has enabled us to derive a set of important  properties, namely: the independence from $\alpha$ of the BRST invariant correlation functions and an exact all order prediction for the longitudinal part of the gluon propagator which agrees with the available lattice numerical simulations as well as with the results based on the analysis of the Dyson-Schwinger equations, see eq.~\eqref{conngp} and related comments at the end of Section \ref{specify}. Recent progress on the extension of the Gribov-Zwanziger set up to the linear covariant gauges was also done in \cite{Moshin:2015gsa,Reshetnyak:2016pkg}. It is worth mentioning that in \cite{Capri:2015pfa}, a non-perturbative BRST symmetry was constructed for the Gribov-Zwanziger action in the maximal Abelian gauge and in \cite{Pereira:2016fpn}, a non-perturbative BRST quantization was proposed for Curci-Ferrari gauges.

The main tool employed in the analysis  \cite{Capri:2015ixa,Capri:2016aqq} has been the introduction of a transverse and order by order gauge invariant field $A^h_\mu$
\begin{equation}
\partial_\mu A^h_\mu=0\;, \qquad \delta A^h_\mu = 0 \;,  \label{gif}
\end{equation}
where $\delta$ stands for the generator of an infinitesimal gauge transformation. As a consequence, the correlation function
\begin{equation}
\langle A^h_\mu(k) A^h_\nu(-k) \rangle  \;, \label{corrfahah}
\end{equation}
is transverse and turns out to be left invariant by the BRST transformations.  As such, it is independent of the gauge parameter $\alpha$ entering the gauge condition. For the benefit of the reader, some details of the construction of the transverse gauge invariant field $A^h_\mu$ have been surveyed in Section \ref{survey}.

The aim of the present work is that of establishing useful relationships between the correlation function \eqref{corrfahah} and the transverse component of the gluon propagator, i.e.~
\begin{equation}
\langle A_\mu(k) A_\nu(-k) \rangle^T = \left( \delta_{\mu \rho} - \frac{k_\mu k_\rho}{k^2} \right) \langle A_\rho(k) A_\nu(-k) \rangle  \;.     \label{gt}
\end{equation}
In particular, we shall be able to show that:
\begin{itemize}
\item the poles of the transverse component of the gluon propagator \eqref{gt} are independent of the gauge parameter $\alpha$. This nice property follows from the Nielsen identities for the two-point gluon correlation function which  can be derived from the Slavnov-Taylor identities corresponding to the exact nilpotent BRST symmetry of the Gribov-Zwanziger action in the linear covariant gauges \cite{Capri:2015ixa,Capri:2016aqq}. We point out that, in the present case, the study of the Nielsen identities requires a lengthy analysis, due to the existence of a nontrivial set of mixed propagators, a structure typical of the local Gribov-Zwanziger formulation. Sections \ref{specify}, \ref{Nielsen} and Appendices \ref{pac}, \ref{apdd} contain the detailed analysis of the structure of the Nielsen identities. We will also briefly discuss the relation between the Nielsen identities and Landau-Khalatnikov-Fradkin transformations.
\item a second property which we shall be able to prove is that the BRST invariant correlation function \eqref{corrfahah} coincides with the gluon propagator evaluated in the Landau gauge, namely
\begin{equation}
\langle A^h_\mu(k) A^h_\nu(-k) \rangle = \langle A_\mu(k) A_\nu(-k) \rangle_{\rm Landau} = \langle A_\mu(k) A_\nu(-k) \rangle_{\alpha=0}  \;, \label{second}
\end{equation}
a relation which gives a quite practical way to evaluate $\langle A^h_\mu(k) A^h_\nu(-k) \rangle$. Moreover, taking into account that the poles of the transverse part of the gluon propagator, eq.~\eqref{gt}, are independent of $\alpha$, it follows immediately that the poles of $\langle A^h_\mu(k) A^h_\nu(-k) \rangle$ and those of   $\langle A_\mu(k) A_\nu(-k) \rangle^T$ are the same, and this for a generic value of $\alpha$.
\item these two properties enable us to consider the BRST invariant correlation function  $\langle A^h_\mu(k) A^h_\nu(-k) \rangle$  as the natural candidate to discuss the positivity violation of the gluon propagator in a BRST and $\alpha$-independent way, via the evaluation of the corresponding temporal Schwinger correlator, a topic which will be addressed in Section \ref{gauge}.  This is a rather relevant issue, as the positivity violation is nowadays taken as a strong indication of gluon confinement, see for instance \cite{Krein:1990sf,Hawes:1993ef,Roberts:1994dr,Alkofer:2003jj,Cucchieri:2004mf,Bashir:2005wt,Bashir:2009fv,Ayala:2012pb,Strauss:2012dg,Bashir:2013zha,Dudal:2013yva,Silva:2013foa,Binosi:2016xxu} and references therein. In this sense, it is certainly worth to have at our disposal a BRST invariant framework to look at it.

\end{itemize}
We end the paper with an application to the study of the masses of the transverse component of the gluon propagator when Higgs fields in the fundamental representation of $SU(2)$ are added to the Gribov-Zwanziger action.

\section{ Survey of the construction of the gauge invariant transverse field  $A_\mu^h$}\label{survey}
The gauge invariant configuration $A_\mu^h$, see Appendix \ref{apb} and \cite{Capri:2015ixa},  is constructed by minimizing  the functional $f_A[u]$ along the gauge
orbit of $A_{\mu }$ \cite{Dell'Antonio:1989jn,vanBaal:1991zw,Zwanziger:1990tn}, namely
\begin{eqnarray}
f_A[u] &\equiv &\min_{\{u\}}\mathrm{Tr}\int d^{4}x\,A_{\mu
}^{u}A_{\mu }^{u}\;,
\nonumber \\
A_{\mu }^{u} &=&u^{\dagger }A_{\mu }u+\frac{i}{g}u^{\dagger }\partial _{\mu
}u\;.  \label{Aminn0}
\end{eqnarray}
In particular, the stationarity condition of the functional \eqref{Aminn0}  gives rise to a non-local transverse
field configuration $A^h_\mu$, $\partial_\mu A^h_\mu=0$, which can be expressed as an infinite series in
the gauge field $A_\mu$, i.e.
\begin{eqnarray}
A_{\mu }^{h} &=&\left( \delta _{\mu \nu }-\frac{\partial _{\mu }\partial
_{\nu }}{\partial ^{2}}\right) \phi _{\nu }\;,  \qquad  \partial_\mu A^h_\mu= 0 \;, \nonumber \\
\phi _{\nu } &=&A_{\nu }-ig\left[ \frac{1}{\partial ^{2}}\partial A,A_{\nu
}\right] +\frac{ig}{2}\left[ \frac{1}{\partial ^{2}}\partial A,\partial
_{\nu }\frac{1}{\partial ^{2}}\partial A\right] +O(A^{3})\;.  \label{min0}
\end{eqnarray}
Remarkably, the configuration $A_{\mu }^{h}$ turns out to be left invariant
by infinitesimal gauge transformations order by order in the gauge
coupling $g$ \cite{Lavelle:1995ty} (see also Appendix \ref{apb}  and the next Section) as
\begin{eqnarray}
\delta A_{\mu }^{h} &=&0\;,  \nonumber \\
\delta A_{\mu } &=&-\partial _{\mu }\omega +ig\left[ A_{\mu },\omega \right]
\;.  \label{gio}
\end{eqnarray}
From expression (\ref{Aminn0}) it follows thus that
\begin{eqnarray}
A_{\min }^{2} &=&\mathrm{Tr}\int d^{4}x\,A_{\mu }^{h}A_{\mu }^{h}\;,  \nonumber \\
&=&\frac{1}{2}\int d^{4}x\left[ A_{\mu }^{a}\left( \delta _{\mu \nu }-\frac{%
\partial _{\mu }\partial _{\nu }}{\partial ^{2}}\right) A_{\nu
}^{a}-gf^{abc}\left( \frac{\partial _{\nu }}{\partial ^{2}}\partial
A^{a}\right) \left( \frac{1}{\partial ^{2}}\partial {A}^{b}\right) A_{\nu
}^{c}\right] \;+O(A^{4})\;.  \label{min1}
\end{eqnarray}
The gauge-invariant nature of expression \eqref{min1} can be made manifest by rewriting it in terms of the  field
strength $F_{\mu \nu }$. In fact, as proven in
\cite{Zwanziger:1990tn}, it turns out that
\begin{eqnarray}
A_{\min }^{2} &=&-\frac{1}{2}\mathrm{Tr}\int d^{4}x\left( F_{\mu
\nu }\frac{1}{D^{2}}F_{\mu \nu }+2i\frac{1}{D^{2}}F_{\lambda \mu
}\left[ \frac{1}{D^{2}}D_{\kappa }F_{\kappa \lambda
},\frac{1}{D^{2}}D_{\nu }F_{\nu \mu }\right] \right.
\nonumber \\
&&-2i\left. \frac{1}{D^{2}}F_{\lambda \mu }\left[ \frac{1}{D^{2}}D_{\kappa
}F_{\kappa \nu },\frac{1}{D^{2}}D_{\nu }F_{\lambda \mu }\right] \right)
+O(F^{4})\;,  \label{zzw}
\end{eqnarray}
from which the gauge invariance becomes apparent. The operator $({D^{2}})^{-1}$ in expression
\eqref{zzw} denotes the inverse of the  covariant Laplacian $D^2=D_\mu D_\mu$ with $D_\mu$ being the
covariant derivative \cite{Zwanziger:1990tn}.

\section{Specification of a local and BRST invariant non-perturbative action and its Slavnov-Taylor identities}\label{specify}
Let us proceed by specifying the non-perturbative local BRST invariant action which will be taken as our starting point.   In order to take into account the non-perturbative effects of the existence of the Gribov copies, we shall make use of the BRST invariant  Gribov-Zwanziger action in linear covariant gauges as recently worked out in \cite{Capri:2015ixa,Capri:2016aqq,Capri:2015nzw,Fiorentini:2016rwx}:
\begin{equation}
S = S_{YM} + S_{FP} + S_{GZ} + S_{\tau}  \;, \label{stact}
\end{equation}
where
\begin{equation}
S_{YM}  = \frac{1}{4} \int d^4x  F^a_{\mu\nu} F^a_{\mu\nu}   \;, \label{ym}
\end{equation}
while  $S_{FP}$ denotes the Faddeev-Popov gauge-fixing in linear covariant gauges, i.e.
\begin{equation}
S_{FP} = \int d^{4}x  \left( \frac{\alpha}{2}\,b^{a}b^{a}
+ib^{a}\,\partial_{\mu}A^{a}_{\mu}
+\bar{c}^{a}\partial_{\mu}D^{ab}_{\mu}(A)c^{b}     \right)   \;, \label{sfp}
\end{equation}
where $\alpha $ is  a non-negative gauge parameter, $b^a$ the Lagrange multiplier and $(c^a, {\bar c}^a)$ the Faddeev-Popov ghosts. The Faddeev-Popov operator is given by
\begin{equation}
{\cal M}^{ab}(A)\bullet = - \delta^{ab} \partial^2\bullet + g f^{abc} \partial_\mu(A^c_\mu\bullet)    \;. \label{fpop2}
\end{equation}
The term $S_{GZ}$ in expression \eqref{stact} stands for the Gribov-Zwanziger action in its local form, as constructed in \cite{Capri:2015ixa,Capri:2016aqq,Capri:2015nzw,Fiorentini:2016rwx}, namely
\begin{equation}\label{lact}
S_{GZ}  = \int d^{4}x  \, \left(
 - \bar\varphi^{ac}_{\nu}{\cal M}^{ab}(A^h)\varphi^{bc}_{\nu}
+\bar\omega^{ac}_{\nu}{\cal M}^{ab}(A^h)\omega^{bc}_{\nu}
+\gamma^{2}g\,f^{abc}(A^h)^{a}_{\mu}(\varphi^{bc}_{\mu}+\bar\varphi^{bc}_{\mu})\, \right)\,,
\end{equation}
where ${\cal M}^{ab}(A^h)$ denotes the gauge invariant counterpart of the Faddeev-Popov operator which, as a consequence of the transversality of the configuration $(A^h)^a_\mu$, reads
\begin{equation}
{\cal M}^{ab}(A^h) = - \delta^{ab} \partial^2 + g f^{abc} (A^h)^c_\mu  \partial_\mu  \;.  \label{fpop}
\end{equation}
Unlike expression \eqref{fpop2}, the operator ${\cal M}^{ab}(A^h)$, eq.~\eqref{fpop},  is Hermitian due to the transverse character of $A^h$.

Following \cite{Capri:2015ixa,Capri:2016aqq,Capri:2015nzw,Fiorentini:2016rwx}, the field $A^{h}_\mu$ can be localised by means of the introduction of an auxiliary Stueckelberg field $\xi^a$, i.e.
\begin{equation}
 A^{h}_{\mu}=(A^{h})^{a}_{\mu}T^{a}=h^{\dagger}A^{a}_{\mu}T^{a}h+\frac{i}{g}\,h^{\dagger}\partial_{\mu}h\,,    \label{st}
\end{equation}
with
\begin{equation}
h=e^{ig\,\xi^{a}T^{a}}  \;.
\label{hxi}
\end{equation}
The local invariance under a gauge transform $u\in SU(N)$ of the field $A^h_\mu$ can now also be appreciated from the transformation prescriptions
\begin{equation}
h\to u^\dagger h\,,\quad\ h^\dagger\to h^\dagger u\,,\quad A_\mu \to u^\dagger A_\mu u + \frac{i}{g}u^\dagger \p_\mu u
\end{equation}
The fields $(\bar\varphi^{ab}_\mu,\varphi^{ab}_\mu)$ are a pair of bosonic fields, while $(\bar\omega^{ab}_\mu,\omega^{ab}_\mu)$ are anti-commuting fields. These fields are employed to cast in local form Zwanziger's horizon term needed to get rid of the zero modes affecting the Faddeev-Popov operator \eqref{fpop2}. The mathematical justification of our construction can be found in \cite{Capri:2015ixa}.

Finally, the term
\begin{equation}
S_{\tau} = \int d^4x\; \tau^{a}\,\partial_{\mu}(A^h)^{a}_{\mu}  \;, \label{stau}
\end{equation}
implements, through the Lagrange multiplier $\tau$, the transversality of the field $A^h$, $\partial_{\mu}(A^h)^{a}_{\mu}=0$, which
can be seen as a  constraint  on the Stueckelberg field. Indeed, if the Stueckelberg field $\xi^a$ is
eliminated through the transversality
constraint $\partial_{\mu}(A^{h})^{a}_{\mu}=0$, we get back the non-local
expression for the field $A^h_\mu$, eq.~\eqref{min0}. This constraint also plays a crucial role to maintain the ultraviolet renormalizability of the theory \cite{Fiorentini:2016rwx,fiorentini2}. If we simply set $S_\tau=0$, we would end up with similar power counting non-renormalizability issues as those plaguing the original Stueckelberg model \cite{Ferrari:2004pd}.

As pointed out in  \cite{Capri:2015ixa,Capri:2016aqq,Capri:2015nzw,Fiorentini:2016rwx}, the action $S$ enjoys an exact nilpotent BRST invariance, namely
\begin{equation}
s S = 0 \;,\;\;\; s^2 =0 \label{ex}
\end{equation}
with the following full set of local transformations defined as
\begin{eqnarray}
s A^{a}_{\mu}&=&-D^{ab}_{\mu}c^{b}\,,\;\;\;\; s c^{a}=\frac{g}{2}f^{abc}c^{b}c^{c}\,,\nonumber\\
s \bar{c}^{a}&=&ib^{a}\,,\;\;\;\;
s b^{a}= 0\,.  \nonumber\\
s h^{ij} &=& -ig c^a (T^a)^{ik} h^{kj}  \; \nonumber\\
s \varphi^{ab}_{\mu}&=& 0 \,,\;\;\;\; s \omega^{ab}_{\mu}=0\,,\nonumber\\
s\bar\omega^{ab}_{\mu}&=&0  \,,\;\;\;\; s\bar\varphi^{ab}_{\mu}=0\,,\nonumber\\
s\tau^{a}&=&0.
 \label{brstgamma}
\end{eqnarray}
The BRST invariance of the action $S$ follows immediately by noticing  that  the field $A^{h}_{\mu}$, eq.~\eqref{st}, is left invariant under the BRST transformations, i.e.
\begin{equation}
s  (A^h)^a_\mu = 0  \;. \label{sah}
\end{equation}
Also, the BRST transformation of the Stueckelberg field  $\xi^a$ can be constructed iteratively from $(s h^{ij})$, obtaining
\begin{equation}
s \xi^a=  - c^a + \frac{g}{2} f^{abc}c^b \xi^c - \frac{g^2}{12} f^{amr} f^{mpq} c^p \xi^q \xi^r + O(g^3)    \;.
\label{eqsxi}
\end{equation}
\subsection{Slavnov-Taylor identities}
The BRST invariance of the action $S$ can be translated at the functional level into powerful Slavnov-Taylor identities.  To that purpose we employ the  trick of extending the BRST transformations on the gauge parameter $\alpha$, see  \cite{Piguet:1995er,Piguet:1984js,Capri:2016aqq}, i.e.
\begin{equation}
s \alpha = \chi \;, \qquad s \chi = 0 \;, \label{extbrst}
\end{equation}
where $\chi$ is a parameter with ghost number 1, which will be set to zero  to restore the initial theory. As explained in  \cite{Piguet:1995er,Piguet:1984js,Capri:2016aqq}, the extended BRST transformations, eqs.~\eqref{brstgamma}, \eqref{extbrst}, will permit us to keep control of the dependence of the Green functions from the gauge parameter $\alpha$ at the quantum level.

Taking into account the extended BRST transformation \eqref{extbrst}, the gauge fixing term becomes now
\begin{equation}
s \int d^4x \left( -i \frac{\alpha}{2} {\bar c}^a b^a + {\bar c}^a \partial_\mu A^a_\mu \right) = \int d^4x \left( \frac{\alpha}{2} b^a b^a + i b^a  \partial_\mu A^a_\mu - i \frac{\chi}{2} {\bar c}^a b^a + {\bar c}^a \partial_\mu D_\mu^{ab}(A) c^b    \right)  \;, \label{ngf}
\end{equation}
so that the action \eqref{stact}  reads
\begin{eqnarray}
S &=& S_{YM}  + \int d^4x \left( \alpha\frac{b^ab^a}{2} + ib^a \partial_\mu A_\mu^a - i \frac{\chi}{2} {\bar c}^a b^a + {\bar c}^a \partial_\mu D_\mu^{ab}(A) c^b \right) +\int d^4x \;\tau^{a}\,\partial_{\mu}(A^h)^{a}_{\mu}
    \;\nonumber\\
&& \hspace{1cm} + \int\,d^4x \left(-\bar\varphi^{ac}_\mu{\cal M}(A^h)^{ab}\varphi^{bc}_\mu + \bar\omega^{ac}_\mu{\cal M}(A^h)^{ab}\omega^{bc}_\mu
+ g {\gamma^2}\,f^{abc}(A^h)^a_\mu(\varphi_\mu^{bc} + \bar\varphi^{bc}_\mu)\right) \;.  \nonumber \\
 \label{gzlocalhorizon2b}
\end{eqnarray}
We are now ready to establish the Ward identities of the theory. Following the general procedure of the algebraic renormalization  \cite{Piguet:1995er}, we introduce a set of BRST-invariant external sources $(\Omega^a_\mu, L^a, K^a, {\cal J}^a_\mu)$ coupled, respectively,  to the non-linear BRST variations of the elementary fields $(A^a_\mu, c^a, \xi^a)$ as well to the composite operator $(A^h)^a_\mu$. Namely, we start with the complete classical action
\begin{equation}
\Sigma=S  + \int d^4x\;{\cal J}^a_\mu (A^{h})^a_\mu + \int d^{4}x \left(\Omega^{a}_{\mu}\,(s A^{a}_{\mu})+L^{a}\,(s c^{a})+K^{a}\,(s\xi^{a} ) \right)    \;, \label{compact}
\end{equation}
where
\begin{equation}
s \Sigma=0\,.
\end{equation}
The complete action $\Sigma$ turns out to obey the following Slavnov-Taylor identity,
\begin{equation}
\mathcal{S}(\Sigma) = 0 \;, \label{stid}
\end{equation}
where
\begin{equation}
\mathcal{S}(\Sigma)=\int d^{4}x \left(
\frac{\delta\Sigma}{\delta\Omega^{a}_{\mu}}\frac{\delta\Sigma}{\delta A^{a}_{\mu}}
+\frac{\delta\Sigma}{\delta L^{a}}\frac{\delta\Sigma}{\delta c^{a}}
+\frac{\delta\Sigma}{\delta K^{a}}\frac{\delta\Sigma}{\delta \xi^{a}}
+ib^{a}\,\frac{\delta\Sigma}{\delta\bar{c}^{a}}  \right)
+\chi\,\frac{\partial\Sigma}{\partial\alpha} \,.    \label{stc}
\end{equation}
It was already shown in \cite{Fiorentini:2016rwx} that, when the Gribov horizon is removed, corresponding to set $\gamma^2=0$, the action $\Sigma$, eq.~\eqref{compact}, is renormalizable to all orders of perturbation theory. Relying on the discussion of \cite{Capri:2015mna}, which was essentially based on the observation that the Gribov-type gluon propagator following from the action $S$, see  eqs.~\eqref{conngp}, \eqref{original}, displays a scalar form factor that can be decomposed as
\begin{equation}\label{uvfall}
\frac{k^2}{k^4+2g^2N\gamma^4}=\frac{1}{k^2}-\frac{2g^2N\gamma^4}{k^2(k^4+2g^2N\gamma^4)}   \;,
\end{equation}
and generalizations thereof, one does expect that, once renormalizability has been proven for $\gamma^2=0$, it will be preserved when $\gamma\neq0$, given the strongly suppressed UV fall-off of the second term in eq.~\eqref{uvfall}, which encodes in fact the dependence from the parameter $\gamma$.  A formal proof to all orders based on the Ward identities is under construction \cite{fiorentini2} and will be presented in a separate detailed work.  Keeping this in mind, the Slavnov-Taylor identities hold at the quantum level, namely
\begin{equation}
\mathcal{S}(\Gamma) = 0 \;, \label{qstid}
\end{equation}
with
\begin{equation}
\mathcal{S}(\Gamma)=\int d^{4}x \left(
\frac{\delta\Gamma}{\delta\Omega^{a}_{\mu}}\frac{\delta\Gamma}{\delta A^{a}_{\mu}}
+\frac{\delta\Gamma}{\delta L^{a}}\frac{\delta\Gamma}{\delta c^{a}}
+\frac{\delta\Gamma}{\delta K^{a}}\frac{\delta\Gamma}{\delta \xi^{a}}
+ib^{a}\,\frac{\delta\Gamma}{\delta\bar{c}^{a}}  \right)
+\chi\,\frac{\partial\Gamma}{\partial\alpha} \,,   \label{qstc}
\end{equation}
where $\Gamma$ denotes the generator of the $1PI$ Green functions of the model. The identities \eqref{qstid} have far-reaching consequences, already exploited in part in  \cite{Capri:2016aqq}, where an all order algebraic proof of the independence from the gauge parameter $\alpha$ of the correlation functions of BRST invariant  operators has been given, together with an exact prediction for the longitudinal part of the gluon propagator.

Let us give a closer look at the two-point correlation functions of the model. To that end we introduce the generator ${\cal Z}^c$ of the connected Green's functions through the Legendre transformation
\begin{equation}
\Gamma = {\cal Z}^c + \sum_{i} \int d^4x J_i(x)  \phi_i(x)    \;, \label{zc}
\end{equation}
\begin{equation}
\phi_i(x)  = - \frac{\delta {\cal Z}^c}{\delta J_i}(x)  \;, \qquad   J_i(x) = \frac{\delta \Gamma}{\delta \phi_i(x)}     \;, \label{f}
\end{equation}
where $\{ \phi_i \}$ is a short-hand notation for all fields  and $\{J_i\}$ for the external sources introduced for each field $\phi_i$. The propagators of the elementary fields  $\langle \phi_i(x) \phi_j(y) \rangle   $, corresponding to the connected two-point correlation functions are given by
\begin{equation}
G_{\phi_{i} \phi_{j} }(x-y) = \langle \phi_i(x) \phi_j(y) \rangle = \frac{  \delta {\cal Z}^c}{\delta J_i(x) \delta J_j(y) } \Big|_{J=0}  \;.
\end{equation}
Also, from eq.~\eqref{f}, we get\footnote{The sum over $k$ implicitly includes an integration.}
\begin{equation}
\delta_{ij} = \frac{ \delta^2 \Gamma}{\delta \phi_i \delta J_j} = \sum_k   \frac{ \delta^2 \Gamma}{\delta \phi_i \delta \phi_k} \frac{\delta \phi_k}{\delta J_j} = -  \sum_k   \frac{ \delta^2 \Gamma}{\delta \phi_i \delta \phi_k} \frac{\delta^2   {\cal Z}^c     }  {\delta J_k \delta J_j}    \;, \label{c1}
\end{equation}
i.e.
\begin{equation}
\sum_k \Gamma_{  {\phi_{i} \phi_{k} } }  G_{\phi_{k} \phi_{j} }   = - \delta_{ij}    \label{dt}
\end{equation}
where we have defined $\Gamma_{\phi_i\phi_k}\equiv \frac{\delta^2\Gamma}{\delta\phi_i\delta\phi_k}$.

When written in terms of the connected generating functional, the Slavnov-Taylor identity \eqref{qstid} takes the form
\begin{equation}
\int d^{4}x \left(
J_{A_\mu^a}(x) \frac{\delta  {\cal Z}^c }{\delta \Omega^{a}_{\mu}(x)}
-J_{c^a} (x) \frac{\delta {\cal Z}^c }{\delta L^{a}(x)}
+J_{\xi^a}(x) \frac{\delta {\cal Z}^c}{\delta K^{a}(x)}
+i  J_{{\bar c}^a} (x) \,\frac{\delta  {\cal Z}^c }{\delta J_{b^a}(x) }  \right)
+\chi\,\frac{\partial  {\cal Z}^c}{\partial\alpha} = 0 \,. \label{qstcz}
\end{equation}
Acting, for example, with the test operators
\begin{equation}
\frac{  \delta^2}{\delta J_{\varphi^{ab}_\mu}(x) \delta J_{\bar{\varphi}^{cd}_\nu}(y)} \;, \qquad \frac{  \delta^2}{\delta J_{\varphi^{ab}_\mu}(x) \delta J_{{\varphi}^{cd}_\nu}(y)} \;, \qquad \frac{  \delta^2}{\delta J_{\bar \varphi^{ab}_\mu}(x) \delta J_{\bar{\varphi}^{cd}_\nu}(y)}   \;, \label{ts1}
\end{equation}
and setting all sources to zero, we immediately get that the propagators $\langle \varphi^{ab}_\mu(k) \bar{\varphi}^{cd}_\nu(-k) \rangle$,  $\langle \varphi^{ab}_\mu(k) {\varphi}^{cd}_\nu(-k) \rangle$, $\langle \bar \varphi^{ab}_\mu(k) \bar{\varphi}^{cd}_\nu(-k) \rangle$ are independent of the gauge parameter $\alpha$, namely
\begin{equation}
\frac{\partial \langle \varphi^{ab}_\mu(k) \bar{\varphi}^{cd}_\nu(-k) \rangle}{\partial \alpha} =  \frac{\partial \langle \varphi^{ab}_\mu(k) {\varphi}^{cd}_\nu(-k) \rangle}{\partial \alpha} = \frac{\partial \langle \bar \varphi^{ab}_\mu(k) \bar{\varphi}^{cd}_\nu(-k) \rangle}{\partial \alpha}   = 0 \;, \label{ia}
\end{equation}
a result which follows by  observing that the fields $(\varphi^{ab}_\mu, \bar \varphi^{ab}_\mu)$ are left invariant by the BRST transformations and, moreover, they interact only with the BRST invariant field  $(A^{h})^a_\mu$ . Likewise, acting with the test  operators
\begin{equation}
\frac{  \delta^2}{\delta {\cal J}^{a}_\mu(x)\delta {\cal J}^{b}_\nu(y)} \;, \qquad \frac{  \delta^2}{\delta {\cal J}^{a}_\mu(x) \delta J_{{\varphi}^{cd}_\nu}(y)} \;, \qquad \frac{  \delta^2}{\delta {\cal J}^{a}_\mu(x) \delta J_{\bar{\varphi}^{cd}_\nu}(y)}   \;, \label{ts2}
\end{equation}
we get
\begin{equation}
\frac{\partial \langle A^{h,a}_\mu(k) A^{h,b}_\nu(-k) \rangle}{\partial \alpha} =  \frac{\partial \langle A^{h,a}_\mu(k) {\varphi}^{cd}_\nu(-k) \rangle}{\partial \alpha} = \frac{\partial \langle A^{h,a}_\mu(k) \bar{\varphi}^{cd}_\nu(-k) \rangle}{\partial \alpha}   = 0 \;. \label{ia2}
\end{equation}
Finally, we remind that, according to \cite{Capri:2015nzw,Capri:2016aqq}, for the gluon propagator we have
\begin{equation}
\langle A^a_\mu(k)  A^b_\nu(-k) \rangle = \delta^{ab} \left( {\delta_{\mu \nu} - \frac{k_\mu k_\nu}{k^2}} \right) G^T_{AA}(k^2) + \delta^{ab} \frac{\alpha}{k^2}  \frac{k_\mu k_\nu}{k^2}  \;,  \label{conngp}
\end{equation}
showing that the introduction  of the Gribov horizon does not modify the longitudinal component which remains equal to its standard  perturbative expression. This result is supported by independent studies of the linear covariant gauge beyond perturbation theory, see \cite{Cucchieri:2009kk,Cucchieri:2011aa,Bicudo:2015rma} for a lattice verification. Dyson-Schwinger equation's studies of the linear covariant gauges \cite{Aguilar:2015nqa,Huber:2015ria,Aguilar:2016ock} automatically incorporate the aforementioned behaviour of the longitudinal component as given by the standard Slavnov-Taylor identity is part of the premisses in this formalism.   Yet another approach to deal with the linear covariant gauge can be found in \cite{Siringo:2014lva,Siringo:2015gia}.

In the present case, the tree level expression for $G^T_{AA}(k^2)$ is given by the Gribov type propagator
\begin{equation}\label{original}
 G^T_{AA}(k^2)\Big|_{\text{tree\; level}} = \frac{k^2}{k^4 +2 g^2N \gamma^4}\,.
\end{equation}

\section{Nielsen identity for the gluon propagator}\label{Nielsen}
We are now ready to derive the Nielsen identity for the gluon propagator \cite{Breckenridge:1994gs,Gambino:1999ai,DelCima:1999gg}.  Roughly speaking, the Nielsen identities are a way to control the gauge parameter dependence of certain correlation functions  and are ultimately a consequence of the BRST invariance \cite{Nielsen:1975fs}.

Though, in the present case the task is not straightforward, due to the existence of mixed propagators. Let us begin by finding the relationship between the transverse component of the gluon propagator  $ G^T_{AA}(k^2) $ and the $1PI$ two-point functions of the elementary fields. From eq.~\eqref{dt}, we have
\begin{eqnarray}\label{dd1}
&& \Gamma_{A^a_\mu A^c_\sigma}(k)  G_{A^c_\sigma A^b_\nu}(-k) + \Gamma_{A^a_\mu b^c}(k) G_{b^c A^b_\nu}(-k) +\Gamma_{A^a_\mu \xi^c}(k) G_{\xi^c A^b_\nu}(-k) + \Gamma_{A^a_\mu \tau^c}(k) G_{\tau^c A^b_\nu}(-k) \nonumber  \\
&& + \Gamma_{A^a_\mu \varphi^{cd}_\sigma}(k) G_{\varphi^{cd}_\sigma A^b_\nu}(-k) + \Gamma_{A^a_\mu \bar \varphi^{cd}_\sigma}(k) G_{\bar \varphi^{cd}_\sigma A^b_\nu}(-k)
=-\delta^{ab} \delta_{\mu \nu}    \;.
\end{eqnarray}
Multiplying by the transverse projector
\begin{equation}
{\cal P}_{\mu \nu}(k) = \delta_{\mu\nu} - \frac{k_\mu k_\nu}{k^2} \;,
\end{equation}
and taking into account Lorentz  invariance, we get
\begin{equation}
 \Gamma^T_{A^a_\mu A^c_\sigma}(k)  G^T_{A^c_\sigma A^b_\nu}(-k)
 +  \Gamma^T_{A^a_\mu  \varphi^{cd}_\sigma}(k) G^T_{ \varphi^{cd}_\sigma A^b_\nu}(-k)  + \Gamma^T_{A^a_\mu \bar \varphi^{cd}_\sigma}(k) G^T_{\bar \varphi^{cd}_\sigma A^b_\nu}(-k)
= -  \delta^{ab}  {\cal P}_{\mu \nu}(k) \;.
\end{equation}
From global color invariance  and the absence of the totally symmetric tensor $d^{abc}$ (cf.~discussion in the next section), we may set\footnote{We shall omit field indices in functional derivatives of $\Gamma$ for notational simplicity.}
\begin{eqnarray}
 \Gamma^T_{A^a_\mu A^c_\sigma}(k)  & = &  \delta^{ac} {\cal P}_{\mu \sigma}(k) \; \Gamma^T_{A A}(k^2)     \;, \nonumber \\
 \Gamma^T_{A^a_\mu  \varphi^{cd}_\sigma}(k) & = &   \Gamma^T_{A^a_\mu  \bar \varphi^{cd}_\sigma}(k) = f^{acd} {\cal P}_{\mu \sigma}(k)  \; \Gamma^T_{A \varphi}(k^2) \;, \nonumber \\
 G^T_{A_\sigma^c A_\nu^b}(-k)&=& \delta^{cb}\mathcal{P}_{\sigma \nu}(k)G^T_{AA}(k^2)\;,\nonumber\\
  G^T_{ \varphi^{cd}_\sigma A^b_\nu}(-k) &=& G^T_{ \bar \varphi^{cd}_\sigma A^b_\nu}(-k) = f^{bcd}  {\cal P}_{\nu \sigma}(k) G^T_{A\varphi}(k^2)  \;, \label{ff}
\end{eqnarray}
so that
\begin{equation}
\Gamma^T_{A A}(k^2) G^T_{A A}(k^2) + 2N  \Gamma^T_{A \varphi}(k^2) G^T_{A  \varphi}(k^2) = - 1   \;, \label{g1}
\end{equation}
which gives
\begin{equation}
\frac{1}{G^T_{A A}} = - \frac{ \Gamma^T_{A A} }{ 1+ 2N  \Gamma^T_{A \varphi} G^T_{A  \varphi}     }       \;. \label{g1a}
\end{equation}
To proceed, let us derive the Nielsen identity for  $\Gamma^T_{A A}$. To that aim, we act on the Slavnov-Taylor identities \eqref{qstid}  with the test operator
\begin{equation}
\frac{\delta^3}{\delta \chi \delta A^a_\mu(x) \delta A^b_\nu(y)}  \; \label{test}
\end{equation}
and set all fields, sources and the parameter $\chi$ to zero. Taking then the Fourier transform, making use of the ghost number conservation,  Lorentz covariance, color invariance, and multiplying everything by the transverse projector ${\cal P}_{\mu \nu}(p)$, one gets
\begin{equation}
\frac{\partial \;  \Gamma^{ T} _{A^a_\mu A^b_\nu}(p^2)  }{\partial \alpha}  = - \Gamma^{ T} _{A^b_\nu A^c_\sigma}(p^2) \; \Gamma^T_{\chi \Omega^c_\sigma A^a_\mu}(p^2)
- \Gamma^{ T} _{A^a_\mu A^c_\sigma}(p^2) \; \Gamma^T_{\chi \Omega^c_\sigma A^b_\nu}(p^2)  \;, \label{n1}
\end{equation}
where $\Gamma^{ T} _{A^a_\mu A^b_\nu}(p^2)$ is the transverse part of the $1PI$ two-point gluon correlation function, i.e.
\begin{equation}
 \Gamma^{T} _{A^a_\mu A^b_\nu}(p^2) = {\cal P}_{\mu \tau} \langle A^a_\tau(p) A^b_\nu(-p)  \rangle_{1PI} = \left(\delta_{\mu\tau} - \frac{p_\mu p_\tau}{p^2}\right) \langle A^a_\tau(p) A^b_\nu(-p)  \rangle_{1PI}    \;, \label{tp}
\end{equation}
and where $\Gamma^T_{\chi \Omega^c_\sigma A^a_\mu}(p^2)$ stands for the Fourier transform of the transverse component of the insertion
\begin{equation}
\frac{\delta^3 \Gamma}{\delta \chi \delta A^a_\mu(x) \delta\Omega^c_\sigma(y)} \Bigg|_{\text{ fields\;$=$\;sources}\;=\; \chi\;=\;0}   \;.  \label{ins}
\end{equation}
Setting now
\begin{equation}
\Gamma^T_{\chi \Omega^c_\sigma A^a_\mu}(p^2) = \delta^{ca} {\cal P}_{\sigma \mu}(p) \; \Gamma^T_{\chi \Omega A}(p^2)  \;, \label{inst}
\end{equation}
eq.~\eqref{n1} becomes
\begin{equation}
\frac{\partial \;  \Gamma^{ T} _{AA} (p^2)  }{\partial \alpha}  =  - 2 \Gamma^{ T} _{AA} (p^2)  \Gamma^T_{\chi \Omega A}(p^2)       \;, \label{nn2}
\end{equation}
expressing the Nielsen identity obeyed by $ \Gamma^{ T} _{AA} (p^2)$.   Likewise, we can derive the Nielsen identity for the mixed $1PI$ form factor $\Gamma^T_{A\varphi}$, eq.~\eqref{ff}, i.e.
\begin{equation}
\frac{\partial \;  \Gamma^{ T} _{A\varphi} (p^2)  }{\partial \alpha}  =  -  \Gamma^{ T} _{AA} (p^2)  \Gamma^T_{\chi \Omega \varphi}(p^2)   - \Gamma^{ T} _{A\varphi} (p^2)  \Gamma^T_{\chi \Omega A}(p^2)    \;, \label{nn2a}
\end{equation}
where $\Gamma^T_{\chi \Omega \varphi}$ stands for the form factor
\begin{equation}
\Gamma^T_{\chi \Omega^d_\sigma \varphi^{bc}_\nu}(p^2) = f^{dbc}  {\cal P}_{\sigma \nu}(p) \; \Gamma^T_{\chi \Omega \varphi}(p^2)  \;, \label{inst2}
\end{equation}
and $\Gamma^T_{\chi \Omega^d_\sigma \varphi^{bc}_\nu}(p^2)$ is  the Fourier transform of the transverse component of the insertion
\begin{equation}
\frac{\delta^3 \Gamma}{\delta \chi \delta \varphi^{bc}_\nu(x)\delta \Omega^d_\sigma(y)} \Bigg|_{\text{ fields\;$=$\;sources\;$=$\; $\chi\;=$\;0}}   \;.  \label{insa}
\end{equation}
We can now derive the Nielsen identity for the gluon propagator $G^T_{A A}(k^2)$. Taking the derivative of eq.~\eqref{g1a}  with respect to the gauge parameter $\alpha$ and making use of eq.~\eqref{nn2}, it turns out that
\begin{equation}
\frac{\partial }{\partial \alpha} \frac{1}{G^T_{AA}}=  - 2\; \frac{ \Gamma^T_{\chi \Omega A}  }{  G^T_{AA} }  - \frac{1}{  G^T_{AA}} \frac{\partial }{\partial \alpha} \log(- G^T_{AA}  \Gamma^{ T} _{AA} )  \;,   \label{n3}
\end{equation}
expressing the Nielsen identity for the transverse component of the gluon propagator $ G^T_{AA}$.

Unfortunately, due to the presence of the term $ \log(- G^T_{AA}  \Gamma^{ T} _{AA} )$, it is yet unclear how eq.~\eqref{n3}  would imply that the poles of $G^{ T} _{AA}$ are $\alpha$-independent. We remind in fact that, in the present case,  $G^T_{AA}  \Gamma^{ T} _{AA}\neq -1$, due to the existence of mixed propagators. A more complicated Nielsen identity involving  the determinant of the $1PI$ two-point function is needed to achieve the desired result, a topic which will be worked out in great detail in the next subsection.
Before delving into those details, let us first spend a few words on the quantity  $\Gamma^T_{\chi \Omega A}$ appearing in eqs.~\eqref{nn2}  and  \eqref{n3}. In particular, looking at expression \eqref{nn2}, one is led to state that the zeroes of $\Gamma^{ T} _{AA} (p^2)$ should be independent of the parameter $\alpha$, due to the presence of  $\Gamma^{ T} _{AA} (p^2)$ itself in the right hand side of eq.~\eqref{nn2}. Evidently, this is true provided the factor $\Gamma^T_{\chi \Omega A}$  is not too singular at the zero of $\Gamma^{ T} _{AA} (p^2)$, so as to compensate the zero itself. This is a not so evident question for which, to our knowledge, no complete answer based on Ward identities is available so far\footnote{We were unable to understand the simple argument provided in \cite[Sect.~4]{Kronfeld:1998di}.}. It is therefore useful to outline some argument in favour of the absence of unwanted singularities in the quantity $\Gamma^T_{\chi \Omega A}$. As shown in Appendix \ref{apdd}, the quantity $\Gamma^T_{\chi \Omega A}$ can be rewritten as
 \begin{equation}
  \Gamma^T_{\chi \Omega A}(p^2) = -\frac{i}{2} \Gamma^T_{AA}(p^2) {\cal G}^T_{(Dc) A}(p^2) - i N \Gamma^T_{A \varphi}(p^2)   {\cal G}^T_{(Dc) \varphi}(p^2)  \;,   \label{ra}
 \end{equation}
where ${\cal G}^T_{(Dc)A}$ and $ {\cal G}^T_{(Dc){\varphi}}$  are the  form factors of the Fourier transform of the connected two-point Green functions $\langle (\int d^4t\; {\bar c}^d(t) b^d(t) ) D^{ae}_{\mu}c^{e}(x)A^{c}_{\sigma}(x_1)\rangle^T_{c}$ and $\langle (\int d^4t\; {\bar c}^d(t) b^d(t) ) (D^{ae}_{\mu}c^{e})(x)  \varphi^{ck}_{\sigma}(x_1) \rangle^T_{c}$. Thus,  the Nielsen identity \eqref{nn2}  becomes
\begin{equation}
\frac{\partial \;  \Gamma^{ T} _{AA}   }{\partial \alpha}  = i \Gamma^{ T} _{AA} \left(    \Gamma^T_{AA} {\cal G}^T_{(Dc) A} +  2 N \Gamma^T_{A \varphi}   {\cal G}^T_{(Dc) \varphi}  \right)         \;, \label{nn2b}
\end{equation}
which turns out to be quite  useful for an order by order Feynman diagrammatic analysis. Let us first focus on the term  $\Gamma^{ T} _{AA}   \Gamma^T_{AA} {\cal G}^T_{(Dc) A} $, which is already present in standard Yang-Mills theory \cite{Breckenridge:1994gs}. In order to have a  compensation at the zero of $\Gamma^{ T} _{AA}$, the connected Green function ${\cal G}^T_{(Dc) A}$ should develop a double pole, which seems  unlikely to happen, at least in an order by order Feynman diagram expansion. This reasoning is also supported by explicit one loop calculations in ordinary Yang-Mills theory \cite{Breckenridge:1994gs}, where the quantity  ${\cal G}^T_{(Dc) A}$  indeed does not develop a double pole. A similar argument applies as well to the second term $\Gamma^{ T} _{AA} \Gamma^T_{A \varphi}   {\cal G}^T_{(Dc) \varphi}$. In summary, in the following we shall assume that the quantity $\Gamma^T_{\chi \Omega A}$ is not too singular to compensate the zeroes of $\Gamma^{ T} _{AA}$. Though, an explicit proof valid to all orders of this statement remains to be worked out, even for standard perturbative QCD.

Before ending this section, it is worth  emphasizing  that the auxiliary fields $(\bar{\varphi}^{ab}_{\mu}, \varphi^{ab}_{\mu}, \bar{\omega}^{ab}_{\mu},  \omega^{ab}_{\mu})$ of the Gribov-Zwanziger action, $S_{GZ}$, eq.~\eqref{lact}, develop their own dynamics, giving rise to additional non-perturbative effects encoded in the formation of BRST invariant dimension two condensates  $\langle A^{h,a}_\mu  A^{h,a}_\mu \rangle$  and $\langle \bar{\varphi}^{ab}_{\mu}\varphi^{ab}_{\mu}-\bar{\omega}^{ab}_{\mu}   \omega^{ab}_{\mu}  \rangle$. As shown in  \cite{Dudal:2007cw,Dudal:2008sp,Dudal:2011gd,Gracey:2010cg}, taking into account the existence of the aforementioned dimension two condensates, leads to a refinement of the Gribov-Zwanziger theory, whose action is given by
\begin{equation}
S_{RGZ} = S_{GZ} + \frac{m^2}{2} \int d^4x\; A^{h,a}_\mu A^{h,a}_\mu - \mu^2 \int d^4x \left( \bar{\varphi}^{ab}_{\mu}\varphi^{ab}_{\mu}-\bar{\omega}^{ab}_{\mu}   \omega^{ab}_{\mu}  \right)   \;, \label{rgz}
\end{equation}
where $S_{GZ}$ is the Gribov-Zwanziger action of eq.~\eqref{lact} and where, as much as the Gribov parameter $\gamma^2$, the parameters $(m^2,\mu^2)$, corresponding to the condensates  $\langle A^{h,a}_\mu A^{h,a}_\mu \rangle$ and $ \langle \bar{\varphi}^{ab}_{\mu}\varphi^{ab}_{\mu}-\bar{\omega}^{ab}_{\mu}   \omega^{ab}_{\mu} \rangle$,  respectively , are determined in a self-consistent way by suitable gap equations \cite{Dudal:2011gd}. Accordingly, the starting action $S$, eq.~\eqref{stact}, gets modified into its refined version
\begin{equation}
S_{R} = S_{YM} + S_{FP} + S_{RGZ} + S_{\tau}  \;, \label{stactref}
\end{equation}
which gives rise to the following tree-level gluon propagator \cite{Capri:2015ixa,Capri:2016aqq,Capri:2015nzw}
\begin{equation}
\langle A^a_\mu(k) A^b_\nu(-k) \rangle_{\rm tree\; level} = \delta^{ab} \frac{k^2 + \mu^2}{(k^2+m^2)(k^2+\mu^2) + 2g^2N \gamma^4} \left( \delta_{\mu\nu} - \frac{k_\mu k_\nu}{k^2} \right) + \frac{\alpha}{k^2}  \frac{k_\mu k_\nu}{k^2} \;. \label{gp}
\end{equation}
One observes that the transverse component of expression \eqref{gp} is suppressed in the infrared region, attaining a  non-vanishing  value at $k=0$, while the longitudinal component still coincides with  the usual perturbative expression of the  linear covariant gauges. Remarkably, expression \eqref{gp} is in  good qualitative agreement with the most recent lattice data on the two-point gluon correlation functions, see \cite{Cucchieri:2009kk,Cucchieri:2011aa,Bicudo:2015rma}.

The BRST invariance and the associated Slavnov-Taylor identities generalize straightforwardly to the refined action $S_{R}$, eq.~\eqref{stactref}. In particular,  the Nielsen identity \eqref{n3} also  holds in the refined case.

\subsection{Nielsen identities for the determinant of the $1PI$ propagator matrix}
It is possible to provide a unifying description of the $\alpha$-dependence of the poles of the mixed propagators.  We depart again from the action in eq.~\eqref{gzlocalhorizon2b}  and set
\begin{equation}\label{det1}
  \left\{\begin{array}{ccc}
    \varphi_\mu^{bc}+\ophi_\mu^{bc} & = & U_\mu^{bc}\, \\
    \varphi_\mu^{bc}-\ophi_\mu^{bc} & = & V_\mu^{bc}   \;.
  \end{array}\right.
\end{equation}
Then it is straightforward to check that in the action \eqref{gzlocalhorizon2b}, we can replace
\begin{eqnarray}\label{det2}
&&-\bar\varphi_\mu^{ac}\mathcal{M}^{ab}(A^h)\varphi_\mu^{bc}+g\gamma^2f^{abc}(A^h)_\mu^a(\varphi_{\mu}^{bc}+\bar\varphi_{\mu}^{bc})\nonumber\\&\to&-U_\mu^{ac}\mathcal{M}^{ab}(A^h)U_\mu^{bc}-V_\mu^{ac}\mathcal{M}^{ab}(A^h)V_\mu^{bc}+g\gamma^2f^{abc}(A^h)_\mu^aU_{\mu}^{bc} \end{eqnarray}
as the residual terms in $\ldots\p A^h$ can be reabsorbed by a harmless shift of the field $\tau$.

For the rest of this  subsection, we can ignore the  fields $V_\mu^{ab}$ as these decouple from the theory. In fact  they do not mix with the gluon field and  can be integrated out exactly together with part of the $(\omega,\bar\omega)$-ghosts.

Next, we shall decompose $U_\mu^{bc}$ into its color symmetric and antisymmetric components, motivated by the presence of the antisymmetric tensor $f^{abc}$ in the  tree level mixing  between the fields $U_\mu^{bc}$ and $(A^h)_\mu^a$. In practice, we set
\begin{eqnarray}
U_\mu^{bc}&=& U_\mu^{[bc]}+U_\mu^{(bc)}\,,
\end{eqnarray}
with
\begin{eqnarray}
 U_\mu^{(bc)}&=& \frac{1}{2}\left(U_\mu^{bc}+U_\mu^{cb}\,,\right)\nonumber\\
U_\mu^{[bc]}&=& \frac{1}{2}\left(U_\mu^{bc}-U_\mu^{cb}\right)\,.
\end{eqnarray}
Clearly, $(A^h)_\mu^a$ mixes only with $U_\mu^{[bc]}$. At tree level, we have $\mathcal{M}^{ab}(A^h)\to -\p^2\delta^{ab}$. As a consequence, there is no apparent mixing between the symmetric and antisymmetric sector. Including interactions, the $\p_\mu\left[f^{abc}(A^h)_\mu^c\bullet\right]$ term in $\mathcal{M}^{ab}(A^h)$ couples the symmetric sector $(bc)$ with the antisymmetric one $[bc]$.  Nonetheless, in what follows we show that one can still exclude that beyond the tree level mixed propagators as $\braket{U^{[ab]}_\mu U^{(bc)}_\nu}_p$ or $\braket{A_\mu^a U^{(cd)}_\nu}_p$ would be nonvanishing. Only $\braket{A_\mu^a U^{[bc]}_\nu}_p$ is relevant.
\begin{itemize}
\item $\braket{A_\mu^a U^{(bc)}_\nu}_p$ should, given the symmetry in $bc$, be proportional to the completely symmetric tensor $d^{abc}$, as the only available independent invariant  rank 3  $SU(N)$ tensors are $f^{abc}$ and $d^{abc}$. Though, given that our theory does not contain vertices in $d^{abc}$, it can never emerge from loop corrections\footnote{See \cite[Sect.~12.4]{Itzykson:1980rh} for a discussion about the tensor $d^{abc}$ and when it can (not) appear.} and as such, $\braket{A_\mu^a U^{(bc)}_\nu}_p\equiv0$ based on global color symmetry.

    \item Global color symmetry can also be invoked to prove that $\braket{U^{[ab]}_\mu U^{(cd)}_\nu}_p\equiv0$. Indeed, given the symmetry properties of $\braket{U^{[ab]}_\mu U^{(cd)}_\nu}_p$, it must be proportional to an invariant rank 4 $SU(N)$ tensor $\mathcal{T}^{abcd}$ which is antisymmetric in $ab$ and symmetric in $cd$. As discussed in \cite{Dittner}, there can only be found 8 independent rank 4 tensors in $SU(N)$. Out of these, only the set
 \begin{equation}
    \mho=\{\delta^{ab}\delta^{cd}, \delta^{ac}\delta^{bd}, \delta^{ad}\delta^{bc}, f^{ace}f^{bde},f^{abe}f^{cde}\}  \;, \label{set}
 \end{equation}
 is relevant in our case, since the other possibilities will either contain the absent $d^{abc}$ tensor, or be completely symmetric in $abcd$ (see also \cite{Gracey:2006dr}). A priori, a potential candidate tensor might be $\mathcal{T}^{abcd}=\text{Tr}\left([t^a,t^b]\{t^b,t^c\}\right)$, but upon closer inspection, $\mathcal{T}^{abcd}\propto f^{abe}d^{cde}$ and, as such, it can again be excluded due to the absence of $d^{abc}$ tensor in the theory. To close the argument, one can check that upon proper (anti-)symmetrization, no tensor $\mathcal{T}^{abcd}$ can be formed with elements of $\mho$.
 \end{itemize}
Thus, having excluded exactly the mixing with $U^{(bc)}$, we can forget about the symmetric sector and focus on the antisymmetric sector. We can further decompose $U_\mu^{[ab]}$ as follows,
\begin{eqnarray}
U_\mu^{[ab]}=\underbrace{\frac{1}{N}f^{abp}f^{pmn}U_\mu^{[mn]}}_{\equiv f^{abp}U_\mu^p}+\underbrace{U_\mu^{[ab]}-\frac{1}{N}f^{abp}f^{pmn}U_\mu^{[mn]}}_{\equiv S_\mu^{[ab]}}\,.
\end{eqnarray}
We notice that $f^{abc}S_\mu^{[ab]}=0$ by using  $f^{abc}f^{dbc}=N\delta^{ad}$. Since $U_\mu^p=\frac{1}{N}f^{pmn}U_\mu^{[mn]}$, the relevant piece of eq.~\eqref{det2} simplifies to
\begin{eqnarray}\label{det3}
\int d^4x\left( \frac{N}{2}U_\mu^a \p^2 U_\mu^a + Ng\gamma^2 A_\mu^a \mathcal{P}_{\mu\nu} U_\nu^a\right)\,.
\end{eqnarray}
Evidently, there will be mixed $(U,A)$ propagators. Thanks to this last decomposition, the color structure of the propagator in the $(U,A)$ sector is drastically simplified, as the only available tensor is now $\delta^{ab}$. Thanks to the orthogonality of $f^{abc}$ and $S_\mu^{[bc]}$, we also get $\braket{A_\mu^a S_\mu^{[bc]}}_p\equiv0$ since the latter can be only proportional to $f^{abc}$.

We are now ready to face the derivation of the Nielsen identities. We reconsider the Slavnov-Taylor identity \eqref{qstc}. After the previous field decomposition, we can derive a similar matrix relation as  in eq.~\eqref{dd1}, viz.
\begin{eqnarray}
&& \Gamma_{A^a_\mu A^c_\sigma}(k)  G_{A^c_\sigma A^b_\nu}(-k) + \Gamma_{A^a_\mu b^c}(k) G_{b^c A^b_\nu}(-k) +\Gamma_{A^a_\mu \xi^c}(k) G_{\xi^c A^b_\nu}(-k) + \Gamma_{A^a_\mu \tau^c}(k) G_{\tau^c A^b_\nu}(-k) \nonumber  \\
&& + \Gamma_{A^a_\mu U^{c}_\sigma}(k) G_{U^{c}_\sigma A^b_\nu}(-k)
=-\delta^{ab} \delta_{\mu \nu}\,.
\end{eqnarray}
As before, without loss of information, we can project the foregoing expression on the transverse subspace, yielding\footnote{To avoid notational clutter, we will refrain from writing the momentum dependence from now on.}
\begin{eqnarray}\label{det3b}
\Gamma^T_{A_\mu^a A_\sigma^c}G^T_{A_\sigma^c A_\nu^b}+\Gamma^T_{A_\mu^a U_\sigma^c}G^T_{U_\sigma^c A_\nu^b}=-\delta^{ab} \mathcal{P}_{\mu \nu}\,.
\end{eqnarray}
From global color invariance and transversality of the ensuing propagators, we get\footnote{At this point, the importance of having reduced the mixing terms to the one between $A_\mu^a$ and $U_\mu^a$ can again be appreciated, otherwise we would have had to parametrize (unrestricted by symmetry) rank 4 propagators as $\braket{\varphi_\mu^{ab}\bar\varphi_\nu^{cd}}$.  }
\begin{eqnarray}
  \left\{\begin{array}{ccl}
    \Gamma^T_{A_\mu^a A_\sigma^c} & = & \delta^{ac} \mathcal{P}_{\mu\sigma}\Gamma_{AA}^T\,, \\
       \Gamma^T_{A_\mu^a U_\sigma^c} & = & \delta^{ac} \mathcal{P}_{\mu\sigma}\Gamma_{AU}^T ~(=~\Gamma^T_{U_\mu^aA_\sigma^c})\,,
  \end{array}\right.
\end{eqnarray}
so that eq.~\eqref{det3b} collapses to
\begin{eqnarray}\label{det4}
\Gamma^T_{AA}G^T_{AA}+\Gamma^T_{AU}G^T_{UA}&=&-1\,.
\end{eqnarray}
Likewise, we can derive that
\begin{eqnarray}\label{det5}
\Gamma^T_{AA}G^T_{AU}+\Gamma^T_{AU}G^T_{UU}&=&0\,,\label{det5a}\\
\Gamma^T_{UA}G^T_{AA}+\Gamma^T_{UU}G^T_{UA}&=&0\,,\label{det5b}\\
\Gamma^T_{UA}G^T_{AU}+\Gamma^T_{UU}G^T_{UU}&=&-1\label{det5c}\,.
\end{eqnarray}
Said otherwise, up to a sign, the matrices
\begin{eqnarray}
\Gamma^T&=&\left(
             \begin{array}{cc}
               \Gamma_{AA}^T & \Gamma_{AU}^T \\
               \Gamma_{AU}^T & \Gamma_{UU}^T \\
             \end{array}
           \right)
\quad\text{and}\quad G^T~=~\left(
             \begin{array}{cc}
               G_{AA}^T & G_{AU}^T \\
               G_{AU}^T & G_{UU}^T \\
             \end{array}
           \right)
\end{eqnarray}
are each other's inverse,
\begin{eqnarray}\label{det6}
\Gamma^T G^T=-1\,.
\end{eqnarray}
Ultimately, we are interested in the poles of $G^T_{AA}$. From eq.~\eqref{det6}, it is clear that the matrix $G^T$, and thus  its   elements, can only develop poles due to zeroes in $\det\Gamma^T$. We do not expect poles at $p^2>0$ in the elements of the $2\times2$ matrix $\Gamma^T$, as these would need to correspond to zeroes in one of the propagators at $p^2>0$. Let us present a justification of this. From eq.~\eqref{det6}, we immediately derive
\begin{eqnarray}\label{dett}
\Gamma^T_{AA}=\frac{G^T_{UU}}{(G_{AU}^{T})^2-G^T_{AA}G^T_{UU}}\,,\quad \Gamma^T_{AU}=-\frac{G^T_{AU}}{(G_{AU}^{T})^2-G^T_{AA}G^T_{UU}}\,,\quad \Gamma^T_{UU}=\frac{G^T_{AA}}{(G_{AU}^{T})^2-G^T_{AA}G^T_{UU}}\,.
\end{eqnarray}
Here, we have taken into account that the matrices are actually symmetric in $(A,U)$.

Assuming that $m_*^2$ is a simple pole of $G^T_{AA}$, i.e.~$G^T_{AA}(m_*^2)=\infty$, we can discriminate between 4 possibilities:
\begin{itemize}
\item $G^T_{AU}(m_*^2)<\infty$ and $G^T_{UU}(m_*^2)<\infty$: it follows that $\Gamma^T_{AA}(m_*^2)=\Gamma^T_{AU}(m_*^2)=0$, while $\Gamma^T_{UU}(m_*^2)=\frac{1}{G^T_{UU}(m_*^2)}<\infty$ to comply with eq.~\eqref{det5c}.
\item $G^T_{AU}(m_*^2)=\infty$ and $G^T_{UU}(m_*^2)<\infty$: in this case, due to the presence of $(G^T_{AU}(m_*^2))^2$ in expressions \eqref{dett}, we get $\Gamma^T_{AA}(m_*^2)=\Gamma^T_{AU}(m_*^2)=\Gamma^T_{UU}(m_*^2)=0$.
\item $G^T_{AU}(m_*^2)<\infty$ and $G^T_{UU}(m_*^2)=\infty$: the relations \eqref{dett} again allow to deduce that $\Gamma^T_{AA}(m_*^2)=\Gamma^T_{AU}(m_*^2)=\Gamma^T_{UU}(m_*^2)=0$.
\item $G^T_{AU}(m_*^2)=\infty$ and $G^T_{UU}(m_*^2)=\infty$: again, we get $\Gamma^T_{AA}(m_*^2)=\Gamma^T_{AU}(m_*^2)=\Gamma^T_{UU}(m_*^2)=0$.
\end{itemize}
So, in all cases the matrix elements of $\Gamma^T$ are nonsingular at the pole $m_*^2 $ of $G_{AA}^T$.

The $\alpha$-dependence of the zeroes of the determinant can now be controlled by a Nielsen identity. In general, we have
\begin{eqnarray}\label{det8}
\frac{\p}{\p \alpha} \det \Gamma^T &=& \det \Gamma^T \text{Tr}\left((\Gamma^T)^{-1}\frac{\p}{\p \alpha}\Gamma^T\right)~=~-\det\Gamma^T \text{Tr}\left(G^T\frac{\p}{\p \alpha}\Gamma^T\right)\,.
\end{eqnarray}
The elements of $\frac{\p}{\p \alpha}\Gamma^T$ correspond to analogous relations as \eqref{nn2}-\eqref{nn2a}. Next to the identity \eqref{nn2}, we also need
\begin{eqnarray}
  \frac{\p \Gamma^T_{AU}}{\p \alpha} &=& -\Gamma_{AA}^T\Gamma^T_{\chi\Omega U}-\Gamma_{AU}^T\Gamma^T_{\chi\Omega A}\,, \label{det7a}\\
  \frac{\p \Gamma^T_{UU}}{\p \alpha} &=& -2\Gamma_{AU}^T\Gamma^T_{\chi\Omega U}\label{det7b}\,.
\end{eqnarray}
The derivation  of the latter equalities goes as usual by acting with the appropriate test operator on the Slavnov-Taylor identity, with similar decompositions as in eq.~\eqref{inst} for the form factors.

We are now armed to compute the $\text{Tr}$ appearing in \eqref{det8}, namely
\begin{eqnarray}
\text{Tr}\left(G^T\frac{\p}{\p\alpha}\Gamma^T\right)&=&G^T_{AA}\frac{\p\Gamma^T_{AA}}{\p\alpha}+2G^T_{AU}\frac{\p\Gamma^T_{AU}}{\p\alpha}+G^T_{UU}\frac{\p\Gamma_{UU}}{\p\alpha}\nonumber\\
&=&-2G^T_{AA}\Gamma^T_{AA}\Gamma^T_{\chi\Omega A}+2G_{AU}^T(-\Gamma^T_{AA}\Gamma_{\chi\Omega U}^T-\Gamma^T_{AU}\Gamma^T_{\chi\Omega A})-2G^T_{UU}\Gamma_{AU}^T\Gamma^T_{\chi\Omega U}\nonumber\\&=&-2\Gamma^T_{\chi\Omega A}
\end{eqnarray}
upon using eqs.~\eqref{det4}-\eqref{det5}.

Eventually, we thus obtain
\begin{eqnarray}\label{det9}
\frac{\p}{\p \alpha} \det \Gamma^T &=& 2(\det \Gamma^T)  \Gamma^T_{\chi\Omega A}\,.
\end{eqnarray}
The fair simplicity of this final expression can be understood from the gauge invariance of the propagator $G^T_{UU}$. We recall that $U_\mu^a$ is a BRST invariant field.
As such, it must hold that $\frac{\p G_{UU}^T}{\p\alpha}=0$. From
\begin{eqnarray}
  G^T_{UU}=-\frac{\Gamma_{AA}^T}{\det \Gamma^T} &\Leftrightarrow&  G^T_{UU}\det \Gamma^T  ~=~ -\Gamma_{AA}^T\,,
\end{eqnarray}
we get
\begin{eqnarray}
  G^T_{UU}\frac{\p}{\p \alpha}\det \Gamma^T &=&-\frac{\p\Gamma_{AA}^T}{\p\alpha}\,,
\end{eqnarray}
or
\begin{eqnarray}
\frac{\p}{\p\alpha}\det \Gamma^T &=& -\frac{1}{G_{UU}^T}\frac{\p\Gamma_{AA}}{\p\alpha}  = \frac{1}{\Gamma^T_{AA}}\det\Gamma^T\frac{\p \Gamma_T^{AA}}{\p\alpha} ~=~2(\det \Gamma^T) \Gamma^T_{\chi\Omega A} \;,
\end{eqnarray}
which implies the desired result that the zeroes of $(\det \Gamma^T )$ are $\alpha$-independent.

It is interesting to mention that the zeroes of the $1PI$ matrix $\Gamma^T$ will in general produce poles in all propagators $G^T_{AA}$, $G^T_{AU}$ and $G^T_{UU}$. As  the poles of the latter  are gauge invariant per construction, this observation already strongly suggests that the  zeroes of $\Gamma^T_{AA}$ will also be gauge invariant, even without using the Nielsen identities. However, we were unable to rule out on general grounds cancellations of possible gauge variant zeroes of $\det \Gamma^T$ with compensating zeroes in the elements of $\Gamma^T$. This necessitated the lengthy Nielsen analysis just presented.

\subsection{Renormalization group invariance of the pole masses}
Another well-known interesting property, next to the gauge parameter independence, is the renormalization group invariance of the pole mass(es).
We first point out that $\det \Gamma^T$ will also obey the renormalization group equation if $\Gamma^T$ does as
\begin{eqnarray}
\mu\frac{d}{d\mu}\det\Gamma^T&=&-\det\Gamma^T\text{Tr}\left(G^T\mu\frac{d}{d\mu}\Gamma^T\right)\,.
\end{eqnarray}
The pole masses were identified as the zeroes of $\det\Gamma$, so working around such zero $m_*^2$, with $\det \Gamma^T=(p^2+m_*^2)R(p^2)$, we immediately get from the foregoing equation that
\begin{eqnarray}
0=\mu\frac{d}{d\mu}\det\Gamma^T&=&R(p^2)\mu\frac{d m_*^2}{d\mu}\,,
\end{eqnarray}
hence
\begin{eqnarray}
\mu\frac{d m_*^2}{d\mu}=0\,.
\end{eqnarray}

\subsection{Removing the Gribov horizon}
Formally, the Gribov horizon can be removed from the theory by setting $\gamma^2=0$, in which case the auxiliary fields can be integrated out, yielding a unity. When $\gamma^2=0$, the action \eqref{rgz} reduces to the BRST invariant massive model studied recently in \cite{Fiorentini:2016rwx}, namely
\begin{equation}\label{dettt}
S_{m} = S_{YM} + S_{FP} + S_{\tau} +  \frac{m^2}{2} \int d^4x\; A^{h,a}_\mu A^{h,a}_\mu \;.
\end{equation}
This model can be regarded as the generalization to the linear covariant gauges of the effective massive model introduced in the Landau gauge in \cite{Tissier:2010ts,Serreau:2012cg,Serreau:2013ila}. Therefore, for $\gamma\to0$, one is back to the case of standard Yang-Mills theory, albeit supplemented with a mass term, leading to
\begin{equation}
\Gamma^{ T(\gamma^2=0)} _{AA}   G^{T (\gamma^2=0)}_{AA}  = - 1 \;, \label{gb1}
\end{equation}
so that eq.~\eqref{n3}  becomes
\begin{equation}
\frac{\partial }{\partial \alpha} \frac{1}{G^{T (\gamma^2=0)}_{AA}} =  - 2\; \frac{ \Gamma^{T(\gamma^2=0)}_{\chi \Omega A}  }{  G^{T (\gamma^2=0)}_{AA} }  \;,   \label{n4}
\end{equation}
which is nothing but the usual Nielsen identity of the standard Yang-Mills theory \cite{Breckenridge:1994gs}.

Let now $m^2_*$ denote the pole of the transverse part of the gluon propagator, i.e.
\begin{equation}
\frac{1}{G^{T (\gamma^2=0)}_{AA}(p^2)} \Bigg|_{p^2=-m^2_*} = 0\;. \label{smp}
\end{equation}
Thus, the Nielsen identity \eqref{n4}  becomes
\begin{equation}
\left( \frac{\partial }{\partial \alpha} \frac{1}{G^{T(\gamma^2=0)}_{AA}} \right)_{p^2=-m^2_*} = 0 \;, \label{nm1}
\end{equation}
implying that the pole mass $m^2_*$ of the transverse component of the gluon propagator $G^{T(\gamma^2=0)}_{AA}$ is independent of the gauge parameter $\alpha$ \cite{Breckenridge:1994gs}. The pole mass in the Landau gauge version of eq.~\eqref{dettt} was studied in \cite{Browne:2004mk,Gracey:2004bk}.

\subsection{Nielsen identities and Landau-Khalatnikov-Fradkin transformations}
As we have just shown for the GZ case, and as it is well known in general, Nielsen identities are a direct consequence of the underlying BRST invariance of the theory and they allow to control the gauge parameter dependence of gauge variant quantities.

There is another class of relations, commonly known as the Landau-Khalatnikov-Fradkin (LKF) transformations, that dictate how to connect $n$-point functions in different gauges \cite{Landau:1955zz,Fradkin:1955jr}. At the level of practical usage, the LKF transformation are usually restricted to the QED fermion propagator, see \cite{Bashir:2002sp,Jia:2016wyu,Pennington:2016vxv} for useful references. Nonetheless, also for the QCD case, some progress has recently been made for the quark propagator up to a certain order in perturbation theory, see \cite{Aslam:2015nia}.

In \cite{Sonoda:2000kn}, it was observed that, at least for the QED case, the LKF transformations can be derived from BRST invariance as well, by introducing an auxiliary Stueckelberg field. This strengthens our intuition that Nielsen identities and LKF transformations should be related in some way, as at the end, both are consequences of BRST invariance. Schematically, a Nielsen identity for a connected two-point function in the absence of mixing looks like
\[\frac{\p}{\p\alpha} G_{\phi \phi} = G_{\phi \phi}  M\]
where $M$ corresponds to the analogue of the composite operator correlation function $\Gamma_{\chi\Omega\phi}$. Then we can write
\[\frac{\p}{\p\alpha}\ln G_{\phi \phi} = M\,,\]
which can be solved for by
\[G_{\phi \phi}^{(\alpha)} = G_{\phi \phi}^{(\alpha=0)}  e^{\int_0^{\alpha} d\alpha' M(\alpha')}\,.  \]
This is an LKF transformation, in the sense that the two-point function at $\alpha$ is given by transforming the two-point function at $\alpha=0$ (Landau gauge) with some suitable form factor $e^{\int_0^{\alpha} M}$.

Let us discuss this here in some more detail for the Abelian case, i.e.~QED, and the photon propagator. For QED, we do not even need to worry about the Gribov problem.
We thus consider the Abelian limit of the extended gauge fixing \eqref{ngf} and add the Dirac action for the fermions, so that
\[ S_{QED} = \int d^4x \left(F_{\mu\nu}^2 + i b\p_\mu A_\mu + \frac{\alpha}{2} b^2 -i\frac{\chi}{2} \bar c b - \bar c \p^2 c+\bar\psi \slashed{D}\psi\right)\,. \]
Considering next the Abelian limit of the STI for the generator $\mathcal{Z}^c$ of connected Green functions, eq.~\eqref{qstcz} and acting on it with the appropriate test operator, we get for the Nielsen identity of the photon propagator
\[\frac{\p}{\p\alpha}\braket{A_\mu A_\nu}_k = k_\mu Z_{\chi c A_\nu}(k^2)\,,\]
where
\[ Z_{\chi c A_\nu}(k^2)  = \Braket{\int (\bar c b) c A_\nu}_k\]
is the Abelian analogue of $\Gamma_{\chi\Omega A}$, but now immediately at the connected level.

In the Abelian case, this form factor $Z_{\chi c A_\nu}(k^2)$ can be computed in a closed form, since $\bar c$, $c$ and $b$ are free fields. Indeed, as $\braket{\bar c c} = \frac{1}{k^2}$ and $\braket{b A_\nu} = \frac{k_\nu}{k^2}$, it follows that
\begin{eqnarray}\label{bas2}
\frac{\p}{\p\alpha}\braket{A_\mu A_\nu}_k = \frac{k_\mu k_\nu}{k^4}\,.
\end{eqnarray}
Indeed, there is only the photon-fermion vertex, so
\[\Braket{\int (\bar c b) c A_\nu}_k=\sum_{n=0}^{\infty}\frac{1}{n!} \Braket{\int (\bar c b) c A_\nu\left[\int (\bar\psi \slashed{A}\psi)\right]^n}\,.\]
If $b$ is contracted with $A_\nu$, we obtain exactly eq.~\eqref{bas2} since we then get a factorisation into $\Braket{\int (\bar c b) c}$ and $\sum_{n=0}^{\infty} \Braket{\int (\bar\psi \slashed{A}\psi)^n}$, but the latter expression equals 1 since it matches to the (photon) propagator~$\times$~self-energy. If we contract $b$ with an $A$ from a vertex, we trivially get zero since this amounts to a contraction between a momentum and a (conserved) Dirac current.

The final resolution of the photon Nielsen identity, eq.~\eqref{bas2}, is nothing else but the photon LKF transformation. The solution to the eq.~\eqref{bas2} is, after integration, exactly given by the LKF relation
\[\braket{A_\mu A_\nu }_k^{(\alpha)} = \braket{A_\mu A_\nu}_k^{(\alpha=0)} + \alpha  \frac{k_\mu k_\nu}{k^4}\,.\]
Of course, in the non-Abelian case, the situation gets more complicated, since the
r.h.s.~of the gluon Nielsen identity depends on $\Gamma_{\chi \Omega A}$, which cannot be evaluated in an exact form anymore. This also means that the corresponding LKF relation, obtainable by integrating the Nielsen identity, can no longer be written in a closed form and one needs to resort to an approximation. This is exactly what is done, to a few orders in perturbation theory, in \cite{Aslam:2015nia} for the quark propagator. We did not consider fermions in our current paper, but needless to say also for those degrees of freedom, a Nielsen identity can be derived. For the standard perturbative result, see for example \cite{Breckenridge:1994gs}. The r.h.s.~of the fermion Nielsen identity will depend on $\Gamma_{\chi \bar K \psi}$ and $\Gamma_{\chi K \bar \psi}$ with $\bar K$ and $K$ the sources coupled to the BRST variations of $\psi$ and $\bar\psi$. Also these can no longer be evaluated in closed form in the QCD case, the QED case was studied in depth in \cite{Jia:2016wyu,Pennington:2016vxv}. We will report on the non-Abelian generalization of the LKF transformations and the link with the Nielsen identities for both gluon and fermion $n$-point functions in more detail elsewhere, with attention for the manifest renormalizability of the construction. Let us end this subsection by mentioning that, in principle, one can also derive LKF transformations for the mixed propagators in the GZ case by integrating the corresponding Nielsen identities, which gives a way to ``move'' from the Landau gauge results to those of a general linear covariant gauge.

\section{The gauge invariant correlation function $\langle A^h_\mu(k) A^h_\nu(-k) \rangle$}\label{gauge}
Having constructed the gauge invariant configuration $A_\mu^h$, we are naturally led to introduce the two-point correlation function
\begin{equation}
\langle A^{h,a}_\mu(k) A^{h,b}_\nu(-k) \rangle = \delta^{ab}  \left( \delta_{\mu\nu} - \frac{k_\mu k_\nu}{k^2} \right)   {\cal D}(k^2)    \;, \label{ahah1}
\end{equation}
which, as a consequence of the transversality of $A^h_\mu$, can be parametrized in terms of a single form factor  ${\cal D}(k^2)$.   Due to the gauge invariance of $A^h_\mu$, the correlation function \eqref{ahah1} is BRST invariant. As such, it has the pleasant property of being independent of the gauge parameter $\alpha$ \cite{Capri:2016aqq}, namely
\begin{equation}
\frac{ \partial {\cal D}(k^2)}{\partial \alpha} = 0 \;. \label{ind}
\end{equation}
Due to its BRST invariant and $\alpha$-independent nature,  the two-point correlation function \eqref{ahah1} can be employed to investigate non-perturbative aspects of the theory. A first important property encoded in the expression \eqref{ahah1} follows from  the following identity
\begin{equation}
\langle A^{h,a}_{\mu}(x)A^{h,b}_{\nu}(y)\rangle = \langle A^{h,a}_{\mu}(x)A^{h,b}_{\nu}(y)\rangle_{\alpha=0} = \langle A^{h,a}_{\mu}(x)A^{h,b}_{\nu}(y)\rangle_{S_{\rm Landau}}   \;, \label{aa1}
\end{equation}
where $S_{\rm Landau}$ is the action  \eqref{stact} in the Landau gauge, i.e.~$\alpha =0$, $\partial_\mu A^a_\mu=0$, namely
\begin{equation}
S_{\rm Landau}  = S_{YM} + S_{{FP}_{\alpha=0} } + S_{GZ}   \;,  \label{stact1}
\end{equation}
with
\begin{equation}
S_{{FP}_{\alpha=0}} = \int d^{4}x  \left( ib^{a}\,\partial_{\mu}A^{a}_{\mu}
+\bar{c}^{a}\partial_{\mu}D^{ab}_{\mu}(A)c^{b}     \right)   \;. \label{sfp1}
\end{equation}
Let us give a closer look at the  correlation function $\langle A^{h,a}_{\mu}(x)A^{h,b}_{\nu}(y)\rangle_{S_{\rm Landau}}$, i.e.
\begin{equation}
\langle A^{h,a}_{\mu}(x)A^{h,b}_{\nu}(y)\rangle_{S_{\rm Landau}} =\frac{\int [D\Phi]\,A^{h,a}_{\mu}(x)A^{h,b}_{\nu}(y)\,e^{-S_{\rm Landau}}}{\int [D\Phi]\,e^{-S_{\rm Landau}}}\,,
\end{equation}
where $[D\Phi]$ is a short hand notation for integration over all fields
\begin{equation}
 [D\Phi] =  DA_\mu D\xi D \varphi_\mu D  {\bar \varphi}_\mu D \omega_\mu D  {\bar \omega}_\mu D b D c D {\bar c}  D{\tau}\;.
\end{equation}
Integrating out the fields $(\tau,b,c,\bar{c})$, we get
\begin{equation}
\langle A^{h,a}_{\mu}(x)A^{h,b}_{\nu}(y)\rangle_{S_{\rm Landau}}=\frac{\int [D{\tilde \Phi}]\,\delta(\partial_{\mu}A^{h}_{\mu})\delta(\partial_{\mu}A_{\mu})\det(-\partial\cdot D)\,A^{h,a}_{\mu}(x)A^{h,b}_{\nu}(y)\,e^{-(S_{YM}+S_{GZ}) }}{\int  [D{\tilde \Phi}] \,\delta(\partial_{\mu}A^{h}_{\mu})\delta(\partial_{\mu}A_{\mu})\det(-\partial\cdot D)\,e^{-(S_{YM}+S_{GZ})}}\,.
\end{equation}
Employing eqs.~\eqref{hh2}, \eqref{phi0} of Appendix \ref{apb}, the equation $\partial_\mu A^h_\mu=0$ can be solved iteratively for $\xi^a$ yielding
\begin{equation}
\xi =\frac{1}{\partial ^{2}}\partial _{\mu }A_{\mu }+i\frac{g}{\partial ^{2}%
}\left[ \partial A,\frac{\partial A}{\partial ^{2}}\right] +i\frac{g}{%
\partial ^{2}}\left[ A_{\mu },\partial _{\mu }\frac{\partial A}{\partial ^{2}%
}\right] +\frac{i}{2}\frac{g}{\partial ^{2}}\left[ \frac{\partial A}{%
\partial ^{2}},\partial A\right] +O(A^{3})\;,  \label{xi1}
\end{equation}
so that we can  eventually  integrate over $\xi^{a}$, obtaining
\begin{equation}
\langle A^{h,a}_{\mu}(x)A^{h,b}_{\nu}(y)\rangle\rangle_{S_{\rm Landau}}=\frac{\int [D{\tilde \Phi}] \,\delta(\partial_{\mu}A_{\mu})\det(-\partial\cdot D)\,A^{h,a}_{\mu}(x)A^{h,b}_{\nu}(y)\,e^{-(S_{YM} + S_{GZ})}}{\int  [D{\tilde \Phi}] \,\delta(\partial_{\mu}A_{\mu})\det(-\partial\cdot D)\,e^{-(S_{YM}+S_{GZ}) }}\,, \label{qa1}
\end{equation}
where $A^h_\mu$ is now given by, see eq.~\eqref{hhh3} of Appendix \ref{apb},
\begin{eqnarray}
A_{\mu }^{h} &=&A_{\mu }-\frac{1}{\partial ^{2}}\partial _{\mu }\partial A-ig%
\frac{\partial _{\mu }}{\partial ^{2}}\left[ A_{\nu },\partial _{\nu }\frac{%
\partial A}{\partial ^{2}}\right] -i\frac{g}{2}\frac{\partial _{\mu }}{%
\partial ^{2}}\left[ \partial A,\frac{1}{\partial ^{2}}\partial A\right]
\nonumber \\
&+&ig\left[ A_{\mu },\frac{1}{\partial ^{2}}\partial A\right] +i\frac{g}{2}%
\left[ \frac{1}{\partial ^{2}}\partial A,\frac{\partial _{\mu }}{\partial
^{2}}\partial A\right] +O(A^{3})\;.  \label{qa2}
\end{eqnarray}
An important remark is in order here. When evaluating $\delta(\p_\mu A_\mu^h)$ by means of the foregoing expression, we tacitly assumed that this is the unique solution making the argument of the $\delta$-function zero. Exactly because the  action $S_{GZ}$ implements the condition ${\cal M}^{ab}(A^h)>0$, we are ensured that there are no other solutions connected via infinitesimal gauge transformations to the $A_\mu^h$ constructed via \eqref{qa2}, as this would require ${\cal M}^{ab}(A^h)$ to have zero modes. This is the best one can achieve in the continuum, as excluding other equivalent field configurations would boil down to knowing how to restrict in Landau gauge to the  fundamental modular region, the region of absolute, rather than local, minima of the functional $f_A[u]$ \cite{Zwanziger:1988jt,Dell'Antonio:1989jn,vanBaal:1991zw}, see Appendix \ref{apb}. Unlike the case of the Gribov region $\Omega$, eq.~\eqref{om},  a local and renormalizable action implementing the restriction to the fundamental modular region is, so far, not at our disposal. In \cite{Zwanziger:2003cf} an argument was given why averages over the Gribov region would coincide with those over the fundamental modular region, but this is an unsettled issue, see \cite{Bornyakov:2013ysa}.

Due to the presence in  eq.~\eqref{qa1} of the delta function $\delta(\partial_{\mu}A_{\mu})$, all terms containing a divergence $\partial A$ vanish, namely
\begin{eqnarray}
\langle A^{h,a}_{\mu}(x)A^{h,b}_{\nu}(y)\rangle_{S_{\rm Landau}}&=&\frac{\int [D{\tilde \Phi}] \,\delta(\partial_{\mu}A_{\mu})\det(-\partial\cdot D)\,A^{h,a}_{\mu}(x)A^{h,b}_{\nu}(y)\,e^{-(S_{YM} + S_{GZ}(A))  }}{\int [D{\tilde \Phi}] \,\delta(\partial_{\mu}A_{\mu})\det(-\partial\cdot D)\,e^{-(S_{YM} + S_{GZ}(A))}}\nonumber\\
&=&\frac{\int [D{\tilde \Phi}]\,\delta(\partial_{\mu}A_{\mu})\det(-\partial\cdot D)\,A^{a}_{\mu}(x)A^{b}_{\nu}(y)\,e^{-(S_{YM}+S_{GZ}(A))}}{\int [D{\tilde \Phi}]\,\delta(\partial_{\mu}A_{\mu})\det(-\partial\cdot D)\,e^{-(S_{YM} + S_{GZ}(A))}}\nonumber\\
&=&\langle A^{a}_{\mu}(x)A^{b}_{\nu}(y)\rangle_{\tilde S}\,,
\end{eqnarray}
where $\tilde S$ stands for the standard Gribov-Zwanziger action in the Landau gauge, namely
\begin{eqnarray}
{\tilde S} & =& S_{YM} + \int d^{4}x  \left( ib^{a}\,\partial_{\mu}A^{a}_{\mu}
+\bar{c}^{a}\partial_{\mu}D^{ab}_{\mu}(A)c^{b}     \right) \nonumber \\
& + &  \int d^{4}x  \, \left(
 - \bar\varphi^{ac}_{\nu}{\cal M}^{ab}(A)\varphi^{bc}_{\nu}
+\bar\omega^{ac}_{\nu}{\cal M}^{ab}(A)\omega^{bc}_{\nu}
+\gamma^{2}g\,f^{abc}(A)^{a}_{\mu}(\varphi^{bc}_{\mu}+\bar\varphi^{bc}_{\mu})\, \right)\,.    \label{ts}
\end{eqnarray}
Finally, we end up with the important result
\begin{equation}
\langle A^{h,a}_{\mu}(x)A^{h,b}_{\nu}(y)\rangle =  \langle A^{a}_{\mu}(x)A^{b}_{\nu}(y)\rangle_{\tilde S}    \;, \label{irr}
\end{equation}
which gives us a practical way of computing the correlator $\langle A^{h,a}_{\mu}(x)A^{h,b}_{\nu}(y)\rangle$. More precisely, the BRST invariant correlation function $\langle A^{h,a}_{\mu}(x)A^{h,b}_{\nu}(y)\rangle$ is obtained by evaluating the  gluon propagator $ \langle A^{a}_{\mu}(x)A^{b}_{\nu}(y)\rangle$ in the Landau gauge with the standard Gribov-Zwanziger action $\tilde S$, eq.~\eqref{ts}.

Furthermore, from eq.~\eqref{irr} and from the previous result on the independence from $\alpha$ of the poles of the transverse part of the gluon propagator  $\langle A^{a}_{\mu}(k)A^{b}_{\nu}(-k)\rangle_{ S}^T={\cal P}_{\mu\tau}(k) \langle A^{a}_{\tau}(k)A^{b}_{\nu}(-k)\rangle_{ S}$ , it immediately follows that the poles of $\langle A^{h,a}_{\mu}(k)A^{h,b}_{\nu}(-k)\rangle$ are precisely those of $\langle A^{a}_{\tau}(k)A^{b}_{\nu}(-k)\rangle_{ S}^T$, providing thus a BRST invariant and $\alpha$-independent way of characterizing the nature of the excitations in the gluon sector within the class of the renormalizable linear covariant gauges.

Another useful quantity which can be introduced by means of expression \eqref{ahah1} is the so called temporal Schwinger correlator ${\cal C}(t)$, defined for $t\geq 0$ as
\begin{equation}
 {\cal C}(t) = \frac{1}{2\pi} \int_{-\infty}^{\infty} dp\; e^{-ipt}   {\cal D}(p^2)     \;, \label{ct}
\end{equation}
which is per construction manifestly BRST invariant and $\alpha$-independent. It is known that the violation of the positivity of the temporal correlator \eqref{ct} is directly related to the impossibility of giving a physical particle interpretation to the BRST invariant correlation function \eqref{ahah1} via a K\"all\'{e}n-Lehmann spectral representation. Suppose in fact that the form factor ${\cal D}(k^2)$ admits a K\"all\'{e}n-Lehmann spectral representation, namely
\begin{equation}
{\cal D}(k^2) = \int_{\tau_0}^\infty d\tau \; \frac{\rho(\tau)}{\tau + k^2} \;, \label{kl}
\end{equation}
where $\rho(\tau)\geq0$ denotes the spectral density and $\tau_0$ the threshold. For the temporal correlator \eqref{ct}, one gets
\begin{equation}
{\cal C}(t) = \frac{1}{2}  \int_{\tau_0}^\infty d\tau\; \frac{e^{-t \sqrt{\tau}}}{\sqrt{\tau}} \rho({\tau})  \;.
\end{equation}
Therefore, if ${\cal C}(t)<0$ for some $t\geq0$, then the spectral density $\rho(\tau)$ cannot be positive everywhere. This implies that the correlation function  \eqref{ahah1} cannot be given a particle interpretation in terms of physical excitations belonging to the spectrum of the theory, a situation which is expected to be physically realized in the confining regime of the theory.  For some more discussion about $\mathcal{C}(t)$, see for instance \cite{Cornwall:2013zra,Lowdon:2015fig}.

It is worth underlining that, actually, the violation of the positivity of the temporal correlator is taken as a strong evidence for gluon confinement, from both analytical and numerical lattice studies of the gluon propagator. We see therefore that the introduction of the correlation function  \eqref{ahah1} enables us to attach a BRST invariant meaning to the positivity violation, through the BRST invariant temporal correlator  \eqref{ct}. Of course, from eq.~\eqref{irr}, it follows
\begin{equation}
{\cal C}(t)_{S} = {\cal C}(t)_{\alpha=0} =  {\cal C}(t)_{\tilde S }\;, \label{ld1}
\end{equation}
giving us a  way of checking the positivity violation. In practice, eq.~\eqref{ld1} tells us that in order to check the positivity violation of the temporal correlator ${\cal C}(t)_{S}$ in the linear covariant gauges, it suffices to look at the temporal correlator ${\cal C}(t)_{\tilde S }$ in the Landau gauge, evaluated with the standard Gribov-Zwanziger action $\tilde S$.

It can be easily checked that positivity is violated for the original Gribov propagator \eqref{original}. A contour integration argument gives
\begin{equation}\label{cc}
\mathcal{C}(t)=\frac{e^{-\frac{\lambda}{\sqrt{2}}t}}{2\lambda}\cos\left(\frac{\pi}{4}+\frac{\lambda}{\sqrt{2}}t\right)\,,
\end{equation}
where we have set $\lambda^4=2g^2N\gamma^4$. Evidently, the r.h.s.~of \eqref{cc} is not positive for all $t$. This was observed before in \cite{Hawes:1993ef}.

Using the same method, a closed expression for ${\cal C}(t)$ can also be obtained for the refined propagator \eqref{gp}, but the final expression is not very instructive to read off the positivity violation with the naked eye. Though, this can be easily checked numerically using lattice input for the dynamical mass scales obtained from fitting expression \eqref{gp}, upon a suitable global rescaling related to a choice of MOM renormalization scale, see \cite{Dudal:2010tf,Cucchieri:2011ig,Oliveira:2012eh,Dudal:2012zx}. The positivity violation can also be directly checked from the lattice viewpoint, either via ${\cal C}(t)$ \cite{Cucchieri:2004mf,Silva:2013foa} or directly from the spectral function \cite{Dudal:2013yva}.

\section{ Application: evaluation of the temporal correlator ${\cal C}(t)$ for $SU(2)$ Yang-Mills-Higgs theory}
Consider the action
\begin{equation}
S = S_{YM}^{Higgs} + S_{FP} + S_{GZ} + S_{\tau}  \;. \label{stacth}
\end{equation}
$S_{YM}^{Higgs}$ stands for the Yang-Mills action in presence of  a Higgs field in the fundamental representation
\begin{equation}
S_{YM}^{Higgs} = \int d^4x \left( \frac{1}{4} F^a_{\mu\nu}F^a_{\mu\nu} + \left( D^{ij}_\mu \phi^j \right)^{\dagger}  \left( D^{ik}_\mu \phi^k \right) +\frac{\lambda}{2} \left(  \phi^{\dagger}\phi - v^2\right)^2  \right)   \;, \label{ymh}
\end{equation}
where
\begin{equation}
D^{ij}_\mu \phi^j = \partial_\mu \phi^i - ig (T^a)^{ij} A^a_\mu \phi^j   \;, \label{cv}
\end{equation}
is the covariant derivative with $\{T^a\}$ being the generators of the gauge group $SU(N)$ in the fundamental representation of the gauge group $SU(N)$, $[T^a,T^b]=if^{abc}T^c$. For simplicity, we will work in the ``freezing'' limit $\lambda\to\infty.$

To discuss the behaviour of the temporal correlator \eqref{ld1} we can make direct use of the results already obtained in \cite{Capri:2012ah}. In particular, according to \cite{Capri:2012ah}, the  propagators  of the theory in the plane $(g,v)$ turn out to be characterised by a separation line $a=1/2$, where $a$ denotes the dimensionless quantity
\begin{equation}\label{a}
a = \frac{g^2 v^2}{4 {\bar \mu}^2 e^{\left( 1 - \frac{32\pi^2}{3g^2} \right)} }  \;,
\end{equation}
where ${\bar \mu}$ is the energy scale of the dimensional regularization in the $\MSbar$ scheme. For generality, we will discuss the propagator and temporal correlator behaviour for all values of the parameter $a$. However, as discussed in \cite{Capri:2012ah}, the analysis leading to the result \eqref{a} can only be trusted for very small or very large values of $a$, related to a balancing between size of the leading logs and coupling constant $g^2$.

Following \cite{Capri:2012ah}, we have the following regions:
\begin{itemize}
\item  for $a > 1/2$, the form factor  ${\cal D}(k^2)$ is of the Yukawa type, i.e.
\begin{equation}
{\cal D}(k^2) = \frac{1}{k^2 + \frac{g^2 v^2}{2}} \;.
\end{equation}
The temporal correlator ${\cal C}(t)_{S}$ is always positive. In this region the usual Higgs mechanism takes place. The BRST invariant correlation function \eqref{ahah1} has a clear and transparent meaning: it describes the three  polarizations of a massive gauge boson characteristic of the Higgs phase . We notice that, for sufficiently weak coupling  $g^2$ and high values of the Higgs VEV $v$, the parameter $a$ will always be bigger than $1/2$, and sufficiently big to trust the leading order analysis presented in \cite{Capri:2012ah}.  Moreover, as pointed out in \cite{Capri:2012ah}, the restriction to the Gribov region in the functional integral is not needed.
\item for $1/e < a <1/2$, the form factor ${\cal D}(k^2)$ turns out to be the sum of two Yukawa terms, namely
\begin{equation}
{\cal D}(k^2) = \frac{{\cal F}_+}{k^2+m^2_+} - \frac{{\cal F}_-}{k^2+m^2_-}  \;, \label{ir}
\end{equation}
where
\begin{eqnarray}
m^2_+ &=& \frac{1}{2} \left(\frac{g^2 v^2}{2} + \sqrt{\frac{g^4 v^4}{4} - \frac{4g^2}{3} \vartheta} \right) \;, \qquad  m^2_- = \frac{1}{2} \left(\frac{g^2 v^2}{2} - \sqrt{\frac{g^4 v^4}{4} - \frac{4g^2}{3} \vartheta} \right)   \nonumber \\
{\cal F}_+ &=&  \frac{m^2_+}{m^2_+-m^2_-}  \;, \qquad {\cal F}_- =  \frac{m^2_-}{m^2_+-m^2_-}    \;, \label{vm}
\end{eqnarray}
and the parameter $\vartheta$ is proportional to the Gribov parameter $\gamma$, being given by the gap equation \cite{Capri:2012ah}
\begin{equation}
\frac{3g^2}{2} \int \frac{d^4q}{(2\pi)^4} \frac{1}{q^4 + \frac{g^2 v^2}{2} q^2 + \frac{g^2}{3} \vartheta} = 1 \;. \label{ge}
\end{equation}
Due to the negative nature of the residue ${\cal F}_-$, we cannot give a physical interpretation to ${\cal D}(k^2)$ in terms of excitations. Though, we underline that, due to the BRST invariant nature of ${\cal D}(k^2)$, the two Yukawa modes corresponding to the masses $m^2_+$ and $m^2_-$ cannot be analysed separately. In other words, BRST invariance requires that the two modes $m^2_+$ and $m^2_-$ belong to a unique, BRST invariant, quantity: ${\cal D}(k^2)$. Let us look thus at the temporal correlator ${\cal C}(t)_{S}$ in this region. A simple calculation gives
\begin{eqnarray}
{\cal C}(t)_{S} & = &  \frac{1}{\pi} \int_{-\infty}^{\infty} dp\; e^{-ipt} \left(  \frac{{\cal F}_+}{p^2+m^2_+} - \frac{{\cal F}_-}{p^2+m^2_-}    \right)     \nonumber \\
& = & \frac{{\cal F}_+}{m_+} e^{-m_+t}\left( 1 - \frac{{\cal F}_-}{{\cal F}_+} \frac{{ m}_+}{{m}_-}  e^{-(m_--m_+)t}  \right)   \nonumber \\
& = & \frac{{\cal F}_+}{m_+} e^{-m_+t}\left( 1 -  \frac{{ m}_-}{{m}_+}  e^{-(m_--m_+)t}  \right)       \label{tci}
\end{eqnarray}
Since $m_-< m_+$, the quantity $\left( 1 -  \frac{{ m}_-}{{m}_+}  e^{-(m_--m_+)t}  \right) $ will become negative for sufficiently large $t$, i.e.
\begin{equation}
t > \frac{1}{m_+ - m_-} \log{\frac{m_+}{m_-} }  \;.
\end{equation}
Therefore, the temporal correlator \eqref{tci} in the intermediate region $1/e < a <1/2$ cannot be given a consistent particle interpretation.  Here, the effects of the Gribov copies start to become relevant, forbidding a particle interpretation of the BRST invariant correlator \eqref{ahah1}.

\item finally, we have the region for $a < 1/e$, in which the two masses $(m^2_+, m^2_-)$ become complex
conjugate and the form factor ${\cal D}(k^2)$ is of the Gribov type, displaying complex poles. Again, in this region, the temporal correlator ${\cal C}(t)_{S}$ becomes negative. As usual, this can be interpreted as the confining sector. This region is realized for sufficiently large values of $g^2$/small values of $v$, thereby corresponding to a strong coupling regime. For sufficiently small $a$, we can again trust the approximation made in \cite{Capri:2012ah}.

Summarizing, we have presented evidence that, for sufficiently small or large values of $a$, there are 2 different regions, with Higgs-like or confining-like properties, which are now identified in a BRST invariant fashion. We are, unfortunately, unable to concretely characterize the possible phase transition between these two different sectors since the ``critical'' values of $a$ are beyond validity of the used expansion. From this perspective, it would be interesting to study the behaviour of the propagator $\braket{A^h A^h}_k$ in a lattice setting.

As our propagator is explicitly BRST invariant, we can try to make a connection with recent works \cite{Maas:2013aia,Maas:2014pba,Torek:2016ede} which introduced a gauge invariant perturbation theory, based on the ideas of \cite{Frohlich:1980gj,Frohlich:1981yi}, see also \cite{Kondo:2016ywd}. Part of the underlying motivation is the Fradkin-Shenker paper \cite{Fradkin:1978dv} which contains a proof that\footnote{At least for the lattice version of the $SU(2)$ gauge-fundamental Higgs model without gauge fixing. We are unaware of any continuum version of their results.} no local observable can discriminate between a Higgs or confining phase. Said otherwise, it is always possible to connect the Higgs and confining ``phase'' in an analytical way. Other part of the motivation is that of constructing the physical spectrum of the theory. For such a goal, one should consider, as appropriate for a gauge theory, gauge invariant bound state operators and identify their poles to get a gauge invariant description of massive gauge bosons, see \cite{Maas:2013aia,Maas:2014pba,Torek:2016ede,Frohlich:1980gj,Frohlich:1981yi}. Coming back to the Fradkin-Shenker result, one might expect that the ensuing spectrum can then ``interpolate'' in an analytical way between the expected Higgs and confined behaviour of the physical degrees of freedom. The gauge invariant perturbation theory comes in about when, after choosing a gauge with associated VEV for the Higgs field\footnote{Giving the Higgs a VEV is a gauge dependent operation. In the lattice formulation, the VEV automatically vanishes. Evidently, this does not mean there can be no Higgs phenomenon, it rather means that the gauge invariant spectrum is not as simply identified as in the gauge variant perturbative setting. }, one can expand the connected two-point  function  of the gauge invariant bound state operator in terms of a two-point function of the standard gauge dependent fields and higher order scattering contributions. In our case, we can do a similar thing, i.e.~we can write, in any gauge,
\begin{equation}\label{mm1}
\braket{A^h A^h}_k= \braket{ A A}^T_k + \braket{\mathcal{O}(A^3)}^T_k
\end{equation}
based on the expansion \eqref{qa1}. The superscript $T$ still means that only the transverse sector is considered. If the BRST invariant l.h.s.~correlation function has (a fortiori gauge invariant) poles, so should the r.h.s.~have, indicating that the (gauge dependent) propagator $\braket{ A A}^T_p$ must also have gauge invariant poles.

\end{itemize}

\section{Conclusion}
In this paper, we have exploited the recently introduced, \cite{Capri:2015ixa,Capri:2016aqq,Capri:2015nzw}, BRST invariant, nonperturbative generalization of the linear covariant gauges, which takes into partial account the Gribov ambiguity which hampers the standard Faddeev-Popov gauge fixing procedure. Thanks to the (local) BRST invariance and ensuing Slavnov-Taylor identity, we were able to derive a set of Nielsen identities for the mixed propagators of this novel Gribov-Zwanziger formulation of the linear covariant gauge, which encompasses the widely studied Landau gauge as a special case. A major result at the level of the transverse form factor of the connected gluon propagator is a proof, based on the Nielsen identities, of the gauge parameter independence of its (complex conjugate) poles. As a byproduct of our analysis, we digressed to some extent how the Landau-Khalatnikov-Fradkin transformations are related to the Nielsen identities.

In addition, we also paid attention to the connected two-point function of a transverse, BRST invariant gluon field, $A_\mu^h$, that enters the formulation. Despite the fact that it corresponds to an infinite series of increasingly nonlocal composite operators, it can be handled with the tools of local quantum field theory after the introduction of suitable auxiliary fields. Its two-point function was shown to be exactly equal to the standard gluon propagator in the Landau gauge, and given the Nielsen identity proof, thereby linking its gauge invariant poles to those of the gluon propagator in any linear covariant gauge. We then used this BRST invariant correlator to study, in a now gauge invariant fashion, the violation of positivity in the gluon sector. Although not a proof of confinement, this is seen by many practitioners in the field as an effective consequence of confinement: there are no observable elementary gluon excitations in the asymptotic $\mathcal{S}$-matrix spectrum.

Of course, next to the elementary gluon degrees of freedom discussed in this paper, QCD also implies confined, or at least unobservable, quarks and ghosts. In forthcoming work, we will extend the tools and results of this paper to the quark and ghost sector. For the quark sector, we have in mind a BRST invariant extension of the preliminary Landau gauge models discussed in \cite{Baulieu:2009xr,Capri:2014bsa,Capri:2014fsa,Dudal:2013vha} that describe a quark propagator with a complex conjugate pole structure, in accordance with lattice fits reported in the literature \cite{Parappilly:2005ei,Furui:2006ks}. In general, a relatively simple parametrization of the quark propagator in terms of pairs of complex conjugate, and thus unphysical, poles has been proven to be rather successful to grasp key features of the QCD spectrum in terms of solutions of the Bethe-Salpeter equations, let us for instance refer to \cite{Roberts:1994dr,Alkofer:2003jj,Alkofer:2000wg,Bhagwat:2002tx,Bashir:2012fs,Eichmann:2016yit}.

One might be worried about the occurrence of complex poles in relation to unitarity, the latter at least for the bound state spectrum of the ``complex'' constituents. This is an open question and certainly deserves further study. That unitarity is not necessarily at odds with complex poles in propagators can be appreciated from \cite{Cutkosky:1969fq}, where an explicit recipe, motivated by the Lee-Wick model \cite{Lee:1969fy,Lee:1970iw} was given to ensure unitarity (and Lorentz covariance), order by order in a Feynman diagrammatic expansion, when pairs of complex conjugate poles are introduced into the theory. More recent applications and insights can be found in \cite{Grinstein:2007mp,Grinstein:2008bg}. This might be an interesting avenue to explore in relation with the complex poles induced by the Gribov-Zwanziger quantization scheme.

In the ghost sector, the Nielsen identities and its possible consequences  may in particular shed some light on the ghost propagator in the linear covariant gauge that for the gauge parameter $\alpha\neq0$ has a quite different behaviour compared to the distinct $\alpha=0$ (Landau gauge) case, at least as reported in \cite{Capri:2015nzw,Aguilar:2015nqa,Huber:2015ria}.

We want to stress that understanding key nonperturbative features of the $n$-point functions of QCD for a general class of gauges is more than just of academic interest. These $n$-point functions are the key ingredients in constructing the QCD spectrum, see \cite{Roberts:1994dr,Alkofer:2000wg,Bhagwat:2002tx,Bashir:2012fs,Eichmann:2016yit}.
So far, such analyses have been restricted to Landau gauge for reasons of simplicity. However, depending on the specific approaches, a lot of modeling, in the form of Ans\"atze for the interaction vertices, are required. Frequently, even the input propagators are modelled with desirable, simplifying forms (see e.g.~\cite{Qin:2011dd} for an overview and relevant references). As such, the true gauge invariant nature of the results becomes clouded, since the Ans\"atze are usually rather specific to Landau gauge and engineered to reproduce certain features of the experimental spectrum.
A truly ab initio computation of the QCD spectrum should display a clean gauge invariant nature, ultimately controlled by the BRST invariance when a gauge fixing is employed. Our work, in addition to that of other approaches as those of \cite{Cucchieri:2009kk,Cucchieri:2011aa,Bicudo:2015rma,Aguilar:2015nqa,Huber:2015ria,Aguilar:2016ock,Siringo:2014lva,Siringo:2015gia} can be seen as a first, small step towards this, as at first one needs to understand the two-point functions.

One final piece of our future effort should be dedicated to a more formal aspect of BRST invariant gauge field theories: to what extent can the new BRST invariance of the current nonperturbative formulation of the linear covariant gauges be used to introduce a well-defined global BRST charge acting on the Hilbert space, and if so, is it still possible to derive a confinement criterion in the sense of ensuring the absence of colored asymptotic states from the physical BRST state cohomology? Almost needless to say, we are referring here to a reanalysis of the Kugo-Ojima confinement criterion \cite{Kugo:1979gm,Kugo:1995km}.

\section*{Acknowledgements.}
We are grateful to A.~Bashir for useful correspondence concerning the LKF transformations. The Conselho Nacional de Desenvolvimento Cient\'{i}fico e
Tecnol\'{o}gico (CNPq-Brazil), the Faperj, Funda{\c{c}}{\~{a}}o de
Amparo {\`{a}} Pesquisa do Estado do Rio de Janeiro, the SR2-UERJ
and the Coordena{\c{c}}{\~{a}}o de Aperfei{\c{c}}oamento de
Pessoal de N{\'\i}vel Superior (CAPES) are gratefully acknowledged
for financial support. S.~P.~Sorella is a level PQ-1 researcher under the program Produtividade em Pesquisa-CNPq, 300698/2009-7;
M.~A.~L.~Capri is a level PQ-2 researcher under the program Produtividade em Pesquisa-CNPq, 307783/2014-6;
A.~D.~Pereira is supported by a postdoctoral fellowship, 150039/2016-6. M.~S.~Guimaraes is supported by the Jovem Cientista do Nosso Estado program - FAPERJ E-26/202.844/2015, is a level PQ-2 researcher under the program Produtividade em Pesquisa-CNPq, 307905/2014-4 and is a Procientista under SR2-UERJ.

\appendix

\section{Properties of the functional $f_{A}[u]$.} \label{apb}In this
 Appendix we recall some useful properties of the functional
$f_{A}[u]$
\begin{equation}
f_{A}[u]\equiv \mathrm{Tr}\int d^{4}x\,A_{\mu }^{u}A_{\mu
}^{u}=\mathrm{Tr}\int d^{4}x\left( u^{\dagger }A_{\mu
}u+\frac{i}{g}u^{\dagger }\partial _{\mu }u\right) \left(
u^{\dagger }A_{\mu }u+\frac{i}{g}u^{\dagger }\partial _{\mu
}u\right) \;. \label{fa}
\end{equation}
For a given gauge field configuration $A_{\mu }$, $f_{A}[u]$ is a functional
defined on the gauge orbit of $A_{\mu }$. Let $\mathcal{A}$ be the space of
connections $A_{\mu }^{a}$ with finite Hilbert norm $||A||$, i.e.
\begin{equation}
||A||^{2}=\mathrm{Tr}\int d^{4}x\,A_{\mu }A{_{\mu }=}\frac{1}{2}\int
d^{4}xA_{\mu }^{a}A_{\mu }^{a}<+\infty \;, \label{norm0}
\end{equation}
and let $\mathcal{U}$ be the space of local gauge transformations $u$ such
that the Hilbert norm $||u^{\dagger }\partial {u}||$ is finite too, namely
\begin{equation}
||u^{\dagger }\partial {u}||^{2}=\mathrm{Tr}\int d^{4}x\,\left(
u^{\dagger }\partial _{\mu }u\right) \left( u^{\dagger }\partial
_{\mu }u\right) <+\infty \;. \label{norm1}
\end{equation}
 As discussed in \cite{Zwanziger:1990tn,Dell'Antonio:1989jn,vanBaal:1991zw,semenov}, the functional $f_{A}[u]$ achieves its absolute minimum on the gauge orbit
of $A_{\mu }$. This proposition means that there exists a $h\in
\mathcal{U}$ such that
\begin{eqnarray}
\delta f_{A}[h] &=&0\;,  \label{impl0} \\
\delta ^{2}f_{A}[h] &\ge &0\;,  \label{impl1} \\
f_{A}[h] &\le &f_{A}[u]\;,\;\;\;\;\;\;\;\forall \,u\in
\mathcal{U}\;. \label{impl2}
\end{eqnarray}
The operator $A_{\min }^{2}$ is thus given by
\begin{equation}
A_{\min }^{2}=\min_{\left\{ u\right\} }\mathrm{Tr}\int
d^{4}x\,A_{\mu }^{u}A_{\mu }^{u}=f_{A}[h]\;.  \label{a2min}
\end{equation}
Let us give a look at the two conditions (\ref{impl0}) and
(\ref{impl1}). To evaluate $\delta f_{A}[h]$ and $\delta
^{2}f_{A}[h]$ we set\footnote{The case of the gauge group $SU(N)$
is considered here.}
\begin{equation}
v=he^{ig\omega }=he^{ig\omega ^{a}T^{a}}\;,  \label{set0}
\end{equation}
\begin{equation}
\left[ T^{a},T^{b}\right] =if^{abc}\;T^c\;,\;\;\;\;\;\mathrm{Tr}\left( T^{a}T^{b}\right) =%
\frac{1}{2}\delta ^{ab}\;,  \label{st000}
\end{equation}
where $\omega $ is an infinitesimal Hermitian matrix and we
compute the linear and quadratic terms of the expansion of the
functional $f_{A}[v]$ in power series of $\omega $. Let us first
obtain an expression for $A_{\mu }^{v}$
\begin{eqnarray}
 A_{\mu }^{v} &=&v^{\dagger }A_{\mu }v+\frac{i}{g}v^{\dagger }\partial _{\mu
}v  =e^{-ig\omega }A_{\mu }^{h}e^{ig\omega }+\frac{i}{g}e^{-ig\omega }\partial
_{\mu }e^{ig\omega }\;.  \label{orbit0}
\end{eqnarray}
To order $\omega ^{2}$,
\begin{eqnarray}
A_{\mu }^{v}
&=&A_{\mu }^{h}+igA_{\mu }^{h}\omega -\frac{g^{2}}{2}A_{\mu }^{h}\omega
^{2}-ig\omega A_{\mu }^{h}+g^{2}\omega A_{\mu }^{h}\omega -\frac{g^{2}}{2}%
\omega ^{2}A_{\mu }^{h}  \nonumber \\
&+&\frac{i}{g}\left( ig\partial _{\mu }\omega -\frac{g^{2}}{2}\left(
\partial _{\mu }\omega \right) \omega -\frac{g^{2}}{2}\omega \partial _{\mu
}\omega +g^{2}\omega \partial _{\mu }\omega \right) +O(\omega ^{3})\;,
\label{ex1}
\end{eqnarray}
so that
\begin{equation}
A_{\mu }^{v}=A_{\mu }^{h}+ig[A_{\mu }^{h},\omega ]+\frac{g^{2}}{2}[[\omega
,A_{\mu }^{h}],\omega ]-\partial _{\mu }\omega +i\frac{g}{2}[\omega
,\partial _{\mu }\omega ]+O(\omega ^{3})\;,  \label{A0}
\end{equation}
A little algebra leads subsequently to
\begin{equation}
f_{A}[v]=f_{A}[h]+2\mathrm{Tr}\int d^{4}x\,\left( \omega \partial
_{\mu }A_{\mu }^{h}\right) -\mathrm{Tr}\int d^{4}x\,\omega
\partial _{\mu }D_{\mu }(A^{h})\omega +O(\omega ^{3})\;,
\label{func2}
\end{equation}
so that
\begin{eqnarray}
\delta f_{A}[h] &=&0\;\;\;\Rightarrow \;\;\;\partial _{\mu }A_{\mu
}^{h}\;=\;0\;,  \nonumber \\
\delta ^{2}f_{A}[h] &>&0\;\;\;\Rightarrow \;\;\;-\partial _{\mu }D{_{\mu }(}%
A^{h}{)}\;>\;0\;.  \label{func3}
\end{eqnarray}
The set of field configurations fulfilling conditions (%
\ref{func3}), i.e.~those defining relative minima of the functional $%
f_{A}[u]$, belong to the Gribov region $\Omega $, with
\begin{equation}
\Omega =\left.\{A_{\mu }\right|\partial _{\mu }A_{\mu
}=0\;\mathrm{and}\;-\partial _{\mu }D_{\mu }(A)>0\}\;.
\label{gribov0}
\end{equation}
Imposing transversality via
$\partial
_{\mu }A_{\mu }^{h}=0$, allows to solve for $h=h(A)$ in a power series in $%
A_{\mu }$. We start from
\begin{equation}
A_{\mu }^{h}=h^{\dagger }A_{\mu }h+\frac{i}{g}h^{\dagger }\partial _{\mu
}h\;,  \label{Ah0}
\end{equation}
with
\begin{equation}
h=e^{ig\phi }=e^{ig\phi ^{a}T^{a}}\;.  \label{h0}
\end{equation}
Let us expand $h$ in powers of $\phi $
\begin{equation}
h=1+ig\phi -\frac{g^{2}}{2}\phi ^{2}+O(\phi ^{3})\;.  \label{hh1}
\end{equation}
From eq.~(\ref{Ah0}) we have
\begin{equation}
A_{\mu }^{h}=A_{\mu }+ig[A_{\mu },\phi ]+g^{2}\phi A_{\mu }\phi -\frac{g^{2}%
}{2}A_{\mu }\phi ^{2}-\frac{g^{2}}{2}\phi ^{2}A_{\mu }-\partial _{\mu }\phi
+i\frac{g}{2}[\phi ,\partial _{\mu }]+O(\phi ^{3})\;.  \label{A1}
\end{equation}
Thus, condition $\partial _{\mu }A_{\mu }^{h}=0$, gives
\begin{eqnarray}
\partial ^{2}\phi &=&\partial _{\mu }A+ig[\partial _{\mu }A_{\mu },\phi
]+ig[A_{\mu },\partial _{\mu }\phi ]+g^{2}\partial _{\mu }\phi A_{\mu }\phi
+g^{2}\phi \partial _{\mu }A_{\mu }\phi +g^{2}\phi A_{\mu }\partial _{\mu
}\phi   \nonumber \\
&-&\frac{g^{2}}{2}\partial _{\mu }A_{\mu }\phi ^{2}-\frac{g^{2}}{2}A_{\mu
}\partial _{\mu }\phi \phi -\frac{g^{2}}{2}A_{\mu }\phi \partial _{\mu }\phi
-\frac{g^{2}}{2}\partial _{\mu }\phi \phi A_{\mu }-\frac{g^{2}}{2}\phi
\partial _{\mu }\phi A_{\mu }-\frac{g^{2}}{2}\phi ^{2}\partial _{\mu }A_{\mu
}  \nonumber \\
&+&i\frac{g}{2}[\phi ,\partial ^{2}\phi ]+O(\phi ^{3})\;.  \label{hh2}
\end{eqnarray}
This equation can be solved iteratively for $\phi $ as a power series in $%
A_{\mu }$,
\begin{equation}
\phi =\frac{1}{\partial ^{2}}\partial _{\mu }A_{\mu }+i\frac{g}{\partial ^{2}%
}\left[ \partial A,\frac{\partial A}{\partial ^{2}}\right] +i\frac{g}{%
\partial ^{2}}\left[ A_{\mu },\partial _{\mu }\frac{\partial A}{\partial ^{2}%
}\right] +\frac{i}{2}\frac{g}{\partial ^{2}}\left[ \frac{\partial A}{%
\partial ^{2}},\partial A\right] +O(A^{3})\;,  \label{phi0}
\end{equation}
which can be simplified to
\begin{eqnarray}
A_{\mu }^{h}
&=&A_{\mu }-\frac{\partial _{\mu }}{\partial ^{2}}\partial A+ig\left[ A_{\mu
},\frac{1}{\partial ^{2}}\partial A\right] +\frac{ig}{2}\left[ \frac{1}{%
\partial ^{2}}\partial A,\partial _{\mu }\frac{1}{\partial ^{2}}\partial
A\right] +ig\frac{\partial _{\mu }}{\partial ^{2}}\left[ \frac{\partial
_{\nu }}{\partial ^{2}}\partial A,A_{\nu }\right]   \nonumber \\
&+&i\frac{g}{2}\frac{\partial _{\mu }}{\partial ^{2}}\left[ \frac{\partial A%
}{\partial ^{2}},\partial A\right] +O(A^{3})\,.  \label{hhh3}
\end{eqnarray}
The transverse field
given in eq.~(\ref {min0}) is, as expected, gauge
invariant.  Let us illustrate this under a gauge transformation
\begin{equation}
\delta A_{\mu }=-\partial _{\mu }\omega +ig[A_{\mu },\omega ]\;.
\label{gauge3}
\end{equation}
Up to the order $O(g^{2})$ we get
\begin{eqnarray}
 \delta \phi _{\nu } &=&-\partial _{\nu }\omega +i\frac{g}{2}\left[ \frac{1}{\partial ^{2}}%
\partial A,\partial _{\nu }\omega \right] +i\frac{g}{2}\left[ \partial _{\nu
}\frac{1}{\partial ^{2}}\partial A,\omega \right] +O(g^{2})\;.  \label{gg2}
\end{eqnarray}
So,
\begin{equation}
\delta \phi _{\nu }=-\partial _{\nu }\left( \omega -i\frac{g}{2}\left[ \frac{%
\partial A}{\partial ^{2}},\omega \right] \right) +O(g^{2})\;,  \label{phi1}
\end{equation}
from which the gauge invariance of $A_{\mu }^{h}$ is established.

\section{Gauge parameter independence of the pole mass of $G^{T}_{AA}$: a slightly different reasoning} \label{pac}
We provide here a second proof of the independence from the gauge parameter $\alpha$ of the poles of the transverse component $G^{T}_{AA}$ of the gluon propagator. In a  quantum field theory which does not have mixed propagators of different fields, eq.~(\ref{dt}) implies, essentially, that the $1PI$ two-point function  is the inverse of the connected two-point function. An immediate consequence of this fact is that the poles of the connected two-point function coincide with the zeroes of the  corresponding $1PI$ two-point function. Therefore, in this simple case, if one is able to prove that the zero/pole of the $1PI$/connected two-point function is independent of the gauge parameter $\alpha$, the independence from the gauge parameter of the pole/zero of the connected/$1PI$ two-point function is a direct consequence.

Nevertheless, as long as  theories with mixed propagators are considered, these properties are lost and one has to be more careful in the analysis of the gauge independence of the poles/zeroes of the connected/$1PI$ two-point functions. In the present case, we are dealing with the Gribov-Zwanziger action which has a large number of fields and of non-trivial mixed propagators, see \cite{Capri:2015ixa,Capri:2016aqq}. However, we were able to derive the identity \eqref{nn2} which tells us that if at $p^2=-m^2$ the two-point function $\Gamma^{T}_{AA}(m^2)$ vanishes and the insertion $\Gamma^{T}_{\chi\Omega A}$ is not too singular, then the zero $m^2$ is independent of $\alpha$, namely,

\begin{equation}
\frac{\partial m^2}{\partial\alpha} = 0\,.
\label{r1}
\end{equation}
Although the $\alpha$-independence of the zero of $\Gamma^{T}_{AA}$ is controlled by \eqref{nn2}, one could be interested on the $\alpha$-independence of the poles of the connected two-point function $G^{T}_{AA}$. For this, let us assume that the pole of $G^{T}_{AA}$ is located at $p^2=-m^2_\ast$ and we split our analysis in two cases:

\begin{enumerate}

\item The pole of $G^{T}_{AA}$ does not coincide with the pole of $G^{T}_{A\varphi}$:

In this case, we consider eq.~(\ref{g1}) at the pole $p^2=-m^{2}_\ast$, namely
\begin{equation}
\Gamma^{T}_{AA}(m^{2}_\ast)G^{T}_{AA}(m^{2}_\ast)+2N\Gamma^{T}_{A\varphi}(m^{2}_\ast)G^{T}_{A\varphi}(m^{2}_\ast)=-1\,.
\label{r2}
\end{equation}
By assumption, $G^{T}_{AA}(m^{2}_{\ast})=\infty$ while $G^{T}_{A\varphi}(m^{2}_{\ast})<\infty$. Since $\Gamma^{T}_{AA}(m^{2}_{\ast})$ and
$\Gamma^{T}_{A\varphi}(m^{2}_{\ast})$ are not singular at the pole $p^2=-m^{2}_\ast $, a property which can be shown in a way  completely similar as done below eq.~\eqref{det6},  the only way the  l.h.s.~of eq.~\eqref{r2} could produce a finite value is by setting $\Gamma^{T}_{AA}(m^2_\ast)=0$. This implies that the pole of $G^{T}_{AA}$ coincides with the zero of $\Gamma^{T}_{AA}$ and by eq.~(\ref{r1}) it is $\alpha$-independent.

\item The pole of $G^{T}_{AA}$ is the same as the pole of $G^{T}_{A\varphi}$:  \label{apd}

For this situation, we consider the following expression
\begin{equation}
\Gamma^{T}_{A^{a}_{\mu}A^{c}_{\lambda}}G^{T}_{A^{c}_{\lambda}\varphi^{be}_{\nu}}(m^2_\ast)+\Gamma^{T}_{A^{a}_{\mu}\varphi^{cd}_{\lambda}}G^{T}_{\varphi^{cd}_{\lambda}\varphi^{be}_{\nu}}(m^2_\ast)+\Gamma^{T}_{A^{a}_{\mu}\bar{\varphi}^{cd}_{\lambda}}G^{T}_{\bar{\varphi}^{cd}_{\lambda}\varphi^{be}_{\nu}}(m^2_\ast)=0\,,
\label{r3}
\end{equation}
which is derived from eq.~(\ref{dt}) by setting $\phi_i = A^{a}_{\mu}$, $\phi_j = \varphi^{be}_{\nu}$ and applying the transverse projector on Lorentz indices. To proceed with the analysis, we subdivide the argument in two cases:

\begin{itemize}

\item The pole of $G^{T}_{A^{c}_{\lambda}\varphi^{be}_{\nu}}$ is the same as the pole of $G^{T}_{\varphi^{cd}_{\lambda}\varphi^{be}_{\nu}}$ and/or the pole of $G^{T}_{\bar{\varphi}^{cd}_{\lambda}\varphi^{be}_{\nu}}$;

As showed in eq.~(\ref{ia}), the two-point functions $G^{T}_{\varphi^{cd}_{\lambda}\varphi^{be}_{\nu}}$ and $G^{T}_{\bar{\varphi}^{cd}_{\lambda}\varphi^{be}_{\nu}}$ are $\alpha$-independent, as a consequence of BRST invariance. As such, their poles are also $\alpha$-independent and, thus, the pole of $G^{T}_{A^{c}_{\lambda}\varphi^{be}_{\nu}}$  is also $\alpha$-independent. By assumption, this pole is the same as the pole of $G^{T}_{AA}$. Hence, the pole of $G^{T}_{AA}$ is independent of $\alpha$.

\item The pole of $G^{T}_{A^{c}_{\lambda}\varphi^{be}_{\nu}}$ is different from the poles of $G^{T}_{\varphi^{cd}_{\lambda}\varphi^{be}_{\nu}}$ and $G^{T}_{\bar{\varphi}^{cd}_{\lambda}\varphi^{be}_{\nu}}$.

In this case, $G^{T}_{\varphi^{cd}_{\lambda}\varphi^{be}_{\nu}}(m^2_\ast)<\infty$ and  $G^{T}_{\bar{\varphi}^{cd}_{\lambda}\varphi^{be}_{\nu}}(m_*^2)<\infty$. Also, $\Gamma^{T}_{A^{a}_{\mu}\varphi^{cd}_{\lambda}}(m^2_\ast)$ and $\Gamma^{T}_{A^{a}_{\mu}\bar{\varphi}^{cd}_{\lambda}}(m^2_\ast)$ are not singular. Since $G^{T}_{A^{c}_{\lambda}\varphi^{be}_{\nu}}(m^2_\ast)= \infty$, the only way the lhs of eq.~(\ref{r3}) can produce a finite value is if $\Gamma^{T}_{A^{a}_{\mu}A^{c}_{\lambda}}(m^2_\ast)=0$. Assuming thus that $\Gamma_{\chi\Omega A}(m^2_\ast)$ is not too singular, we conclude that $m^2_\ast$ is $\alpha$-independent.

\end{itemize}
\end{enumerate}
In summary, if the zeroes of $\Gamma^{T}_{AA}$ are gauge parameter independent then the poles of $G^{T}_{AA}$ also are.

\section{ The insertion $\Gamma^{T}_{\chi\Omega A}$}   \label{apdd}

As already underlined, the Nielsen identity (\ref{nn2}) ensures the gauge parameter independence of the zeroes of $\Gamma^{T}_{AA}$ if the insertion $\Gamma^{T}_{\chi\Omega A}$ is not too singular at the zero. In this  Appendix we work out an expression for such insertion in terms of connected Green functions which turns out to be quite helpful for investigating the nature of $\Gamma^{T}_{AA}$.

To begin with, we write the insertion as
\begin{equation}
\Gamma_{\chi\Omega^{a}_{\mu}A^{b}_{\nu}}=\frac{\partial}{\partial\chi}\frac{\delta}{\delta\Omega^{a}_{\mu}(x)}\frac{\delta}{\delta A^{a}_{\nu}(y)}\Gamma\,,
\label{ins1}
\end{equation}
and we have to act with the transverse projector on eq.~(\ref{ins1}). From eq.~(\ref{zc}), we write
\begin{equation}
\Gamma_{\chi\Omega^{a}_{\mu}A^{b}_{\nu}}=\frac{\partial}{\partial\chi}\frac{\delta}{\delta\Omega^{a}_{\mu}(x)}\frac{\delta}{\delta A^{a}_{\nu}(y)}\left(\mathcal{Z}^{c}+\int d^4x_1~J_i\phi_i\right)=\frac{\partial}{\partial\chi}\frac{\delta}{\delta\Omega^{a}_{\mu}(x)}\frac{\delta\mathcal{Z}^{c}}{\delta A^{a}_{\nu}(y)}\,.
\label{ins2}
\end{equation}
Applying the functional chain rule, we obtain
\begin{eqnarray}
\Gamma_{\chi\Omega^{a}_{\mu}A^{b}_{\nu}}&=&\frac{\partial}{\partial\chi}\frac{\delta}{\delta\Omega^{a}_{\mu}(x)}\left(\int d^4x_1\sum_i \frac{\delta J_i(x_1)}{\delta A^{b}_{\nu}(y)}\frac{\delta\mathcal{Z}^{c}}{\delta J_i(x_1)}\right)\nonumber\\
&=&\int d^4x_1\sum_i\left(\frac{\delta^3J_i(x_1)}{\delta \chi\delta\Omega^{a}_{\mu}(x)\delta A^{b}_{\nu}(y)}\frac{\delta\mathcal{Z}^{c}}{\delta J_i(x_1)}-\frac{\delta^{2}J_i(x_1)}{\delta\Omega^{a}_{\mu}(x)\delta A^{b}_{\nu}(y)}\frac{\delta^2\mathcal{Z}^{c}}{\delta\chi\delta J_i(x_1)}\right.\nonumber\\
&+&\left.\frac{\delta^2 J_i(x_1)}{\delta\chi \delta A^{b}_{\nu}(y)}\frac{\delta^{2}\mathcal{Z}^{c}}{\delta\Omega^{a}_{\mu}(x)\delta J_i (x_1)}+\frac{\delta J_i(x_1)}{\delta A^{b}_{\nu}(y)}\frac{\delta^3\mathcal{Z}^{c}}{\delta\chi\delta\Omega^{a}_{\mu}(x)\delta J_i(x_1)}\right)\,.
\label{ins3}
\end{eqnarray}
Applying the the transverse projector ${\cal P}_{\mu \nu}(p) = \left( \delta_{\mu\nu} - \frac{p_\mu p_\nu}{p^2} \right)$ and taking into account color invariance and ghost number conservation, expression (\ref{ins3}) reduces to
\begin{eqnarray}
\Gamma^{T}_{\chi\Omega^{a}_{\mu}A^{b}_{\nu}}&=&-\frac{i}{2}\int d^4x_1d^4x_2\left[\Gamma^{T}_{A^{b}_{\nu}(y)A^{c}_{\sigma}(x_1)}\langle\bar{c}^{d}_{x_2}b^d_{x_2}D^{ae}_{\mu}c^{e}(x)A^{c}_{\sigma}(x_1)\rangle^T_{c}\right.\nonumber\\
&+&\left.2\Gamma^{T}_{A^{b}_{\nu}(y)\varphi^{ck}_{\sigma}(x_1)}\langle\bar{c}^{d}_{x_2}b^d_{x_2}D^{ae}_{\mu}c^{e}(x)\varphi^{ck}_{\sigma}(x_1)\rangle^T_{c}\right]\,,
\label{ins4}
\end{eqnarray}
with $\langle\ldots\rangle_c$ denoting the connected correlation functions. Passing to Fourier space gives
\begin{equation}
\Gamma^{T}_{\chi\Omega^{a}_{\mu}A^{b}_{\nu}}(p) = -\frac{i}{2} \Gamma^{T}_{A^{b}_{\nu}A^{c}_{\sigma}}(p) {\cal G}^T_{(D_\mu^{ae}c^e)A^c_\sigma}(-p)  - i \Gamma^{T}_{A^{b} _{\nu}\varphi^{ck}_{\sigma}}(p) {\cal G}^T_{(D_\mu^{ae}c^e){\varphi}^{ck}_\sigma}(-p)  \;,
\end{equation}
where ${\cal G}^T_{(D_\mu^{ae}c^e)A^c_\sigma}(p)$ and $ {\cal G}^T_{(D_\mu^{ae}c^e){\varphi}^{ck}_\sigma}(p)$  are  the Fourier transformations of the transverse components of the connected Green functions $\langle (\int d^4t\; {\bar c}^d(t) b^d(t) ) D^{ae}_{\mu}c^{e}(x)A^{c}_{\sigma}(x_1)\rangle^T_{c}$ and  $\langle (\int d^4t\; {\bar c}^d(t) b^d(t) ) D^{ae}_{\mu}c^{e}(x)  \varphi^{ck}_{\sigma}(x_1) \rangle^T_{c}$. From  the decompositions
\begin{eqnarray}
\Gamma^{T}_{\chi\Omega^{a}_{\mu}A^{b}_{\nu}}(p) & = &  \delta^{ab} {\cal P}_{\mu \nu}(p) \Gamma^T_{\chi \Omega A}(p^2)  \;,   \nonumber \\
 \Gamma^{T}_{A^{b}_{\nu}A^{c}_{\sigma}}(p) & = & \delta^{bc} {\cal P}_{ \nu \sigma}(p) \Gamma^T_{AA}(p^2) \;,   \nonumber \\
 \Gamma^{T}_{A^{b} _{\nu}\varphi^{ck}_{\sigma}}(p) & = & f^{bck}  {\cal P}_{ \nu \sigma}(p) \Gamma^T_{A \varphi}(p^2) \;, \nonumber \\
 {\cal G}^T_{(D_\mu^{ae}c^e)A^c_\sigma}(p) & = & \delta^{ac} {\cal P}_{ \mu \sigma}(p) {\cal G}^T_{(Dc) A}(p^2) \;, \nonumber \\
 {\cal G}^T_{(D_\mu^{ae}c^e){\varphi}^{ck}_\sigma}(p) & = & f^{ack} {\cal P}_{ \mu \sigma}(p)  {\cal G}^T_{(Dc) \varphi}(p^2)    \;, \label{rels}
 \end{eqnarray}
 eq.~\eqref{ins4} becomes
 \begin{equation}
  \Gamma^T_{\chi \Omega A}(p^2) = -\frac{i}{2} \Gamma^T_{AA}(p^2) {\cal G}^T_{(Dc) A}(p^2) - i N \Gamma^T_{A \varphi}(p^2)   {\cal G}^T_{(Dc) \varphi}(p^2)  \;,  \label{relf}
 \end{equation}
 which is useful for a better understanding of the Nielsen identity (\ref{nn2}).

\end{document}